\documentclass[a4paper,11pt]{article}
\usepackage{jcappub} 
\usepackage{xcolor}

\newcommand{\LCDM}{${\Lambda}$CDM}
\newcommand{\mbarLCDM}{$\bar{m}{\tilde\Lambda}$CDM}
\newcommand{\mbarEDE}{$\bar{m}$EDE}

\title{\boldmath Vacuum energy interacting with dark matter and dark radiation: alleviating the $H_0$--$S_8$ seesaw}

\author[a]{Li-Yang Gao,}
\affiliation[a]{School of Science, Jiangxi University of Science and Technology, Ganzhou 341000, China}
\author[b,c,d,e]{She-Sheng Xue,}
\affiliation[b]{ICRANet, Piazzale della Repubblica, 10-65122, Pescara, Italy}%
\affiliation[c]{ ICTP-AP, University of Chinese Academy of Sciences, Beijing, China}%
\affiliation[d]{Physics Department, Sapienza University of Rome, P.le A. Moro 5, 00185, Rome, Italy}%
\affiliation[e]{INFN, Sezione di Perugia, Via A. Pascoli, I-06123, Perugia, Italy}%
\author[f, g, h, 1]{Xin Zhang\note{Corresponding author.}}
\affiliation[f]{Liaoning Key Laboratory of Cosmology and Astrophysics,
College of Sciences, Northeastern University, Shenyang 110819, China}%
\affiliation[g]{MOE Key Laboratory of Data Analytics and Optimization for Smart Industry, Northeastern University, Shenyang 110819, China}%
\affiliation[h]{National Frontiers Science Center for Industrial Intelligence and Systems Optimization, Northeastern University, Shenyang 110819, China}%
\emailAdd{liyanggao@jxust.edu.cn, shesheng.xue@gmail.com, zhangxin@mail.neu.edu.cn}

\abstract{Extensions of \LCDM{} that raise~$H_0$ with additional energy before recombination, such as dark radiation and early dark energy (EDE), also raise~$S_8$.
We study whether a late-time energy exchange between the vacuum and the dark sector can offset this increase.
The coupling rate scales as~$\bar m H^{\alpha}$ and is negligible before recombination for~$\alpha \le -2$; the energy is transferred along the four-velocity of geodesic cold dark matter, so the dark matter produced by the vacuum carries no density perturbation and slows the growth of structure.
We add the exchange to dark radiation (\mbarLCDM) and to a one-parameter EDE model (\mbarEDE), and constrain the models with Planck 2018, DESI DR2 BAO and Pantheon+ data (CBS), and with the SH0ES calibration added (CBSH) to test the exchange when~$H_0$ is held at a high value.
For CBS the posteriors favour an energy flow from the vacuum into dark matter at~$1.8$--$2.7\sigma$, and the exchange lowers~$S_8$ by~$0.02$--$0.08$ but does not raise~$H_0$: \mbarLCDM{} gives~$H_0 = 68.83 \pm 0.91~\mathrm{km\,s^{-1}\,Mpc^{-1}}$, $3.1\sigma$ below SH0ES.
For CBSH the exchange removes the increase of~$S_8$ caused by dark radiation, giving~$S_8 = 0.800 \pm 0.016$ at~$H_0 = 71.34 \pm 0.70~\mathrm{km\,s^{-1}\,Mpc^{-1}}$, and about a third of that caused by EDE for~$\alpha = -2$, with the coupling at~$1.1$--$2.0\sigma$.
The reduction of~$S_8$ is concentrated at~$z \lesssim 0.3$: at the redshifts probed by KiDS-Legacy the lensing amplitude decreases by only~$1$--$2\%$, the shear~$\chi^2$ of the dark-radiation models changes by less than~$3$, and the exchange reduces but does not remove the excess shear~$\chi^2$ of EDE.
A late-time exchange can therefore offset part of the~$S_8$ increase of early-time solutions, but it does not ease the Hubble tension, and current data do not require it.}


\begin{document}
\maketitle
\raggedbottom
\section{Introduction}
\label{sec:intro}
The $\Lambda$ cold dark matter (\LCDM{}) model fits a wide range of cosmological observations, including the anisotropies of the cosmic microwave background (CMB)~\cite{Planck:2018vyg}, baryon acoustic oscillations (BAO)~\cite{DESI:2025zgx} and type Ia supernovae (SNe~Ia)~\cite{Scolnic:2021amr,Brout:2022vxf}.
As the precision of these probes has improved, however, several discrepancies have emerged between parameters inferred from the early and the late Universe~\cite{Verde:2019ivm,DiValentino:2021izs,Abdalla:2022yfr,Li:2026asg}.
The most significant is the Hubble tension: within \LCDM, Planck 2018 CMB data give~$H_0 = 67.36 \pm 0.54~\mathrm{km\,s^{-1}\,Mpc^{-1}}$~\cite{Planck:2018vyg}, whereas the SH0ES distance ladder gives~$H_0 = 73.04 \pm 1.04~\mathrm{km\,s^{-1}\,Mpc^{-1}}$~\cite{Riess:2021jrx}, a difference of~$4.8\sigma$.
Observations with the James Webb Space Telescope (JWST) have confirmed the Cepheid calibration~\cite{Riess:2024ohe,Riess:2024vfa}, although other primary distance indicators give~$H_0 \approx 70~\mathrm{km\,s^{-1}\,Mpc^{-1}}$~\cite{Freedman:2024eph}.
A second, weaker discrepancy concerns the amplitude of matter clustering, $S_8 \equiv \sigma_8\sqrt{\Omega_m/0.3}$.
The cosmic-shear analyses of KiDS-1000 and DES~Y3 find~$S_8 = 0.759^{+0.024}_{-0.021}$~\cite{KiDS:2020suj} and~$0.759^{+0.025}_{-0.023}$~\cite{DES:2021bvc,DES:2021vln}, below the Planck \LCDM{} value~$0.832 \pm 0.013$~\cite{Planck:2018vyg} by~$3\sigma$ and~$2.3\sigma$ as quoted in these analyses, whereas the final KiDS-Legacy analysis finds~$S_8 = 0.815^{+0.016}_{-0.021}$, consistent with the CMB~\cite{Wright:2025xka}.
In either case, low-redshift structure probes do not favour an amplitude higher than that of Planck \LCDM.

Many extensions of \LCDM{} have been proposed to raise~$H_0$~\cite{Knox:2019rjx,Schoneberg:2021qvd,DiValentino:2021izs,Guo:2018ans}.
Some of them act before recombination: a higher expansion rate reduces the sound horizon~$r_s$, and a larger~$H_0$ is then needed to keep the angular acoustic scale of the CMB fixed.
Two examples are additional dark radiation (DR), which raises the effective number of relativistic species~$N_{\rm eff}$~\cite{Bernal:2016gxb,Guo:2018ans,Zhao:2017urm,Feng:2017usu,Blinov:2020hmc,Aloni:2021eaq}, and early dark energy (EDE), a scalar field that contributes briefly to the energy density shortly before recombination~\cite{Poulin:2018cxd,Smith:2019ihp,Poulin:2023lkg}.
In both cases the CMB requires a larger physical dark matter density~$\omega_{\rm cdm}$ (and, for EDE, a larger~$n_s$) to compensate for the extra early expansion, so a higher~$H_0$ comes with a higher~$S_8$~\cite{Hill:2020osr,Ivanov:2020ril}.
Easing the Hubble tension in this way therefore worsens the agreement with low-redshift structure probes, a problem referred to as the $H_0$--$S_8$ seesaw~\cite{Allali:2021azp}.
Late-time modifications, such as dynamical or interacting dark energy~\cite{Chevallier:2000qy,Linder:2002et,Wang:2016lxa,DiValentino:2019ffd,Li:2024qso,Li:2025owk,Li:2026xaz,Li:2025muv,Li:2025vuh,Zhang:2026sci}, leave~$r_s$ unchanged, but BAO and SNe~Ia data tightly constrain the low-redshift expansion history and limit how much these models can raise~$H_0$~\cite{Bernal:2016gxb,Aylor:2018drw,Schoneberg:2021qvd}.
An energy exchange between dark energy and dark matter can lower~$S_8$~\cite{Pourtsidou:2016ico,Lucca:2021dxo}, and a coupling between EDE and dark matter has been proposed to offset the increase of~$S_8$ in EDE~\cite{McDonough:2021pdg}.
Reference~\cite{Allali:2021azp} eased the seesaw with a dark sector of two components: a decaying dark-energy fluid that raises~$H_0$ and an ultralight axion that suppresses the matter power spectrum.

In this work we study whether an energy exchange between the vacuum and cold dark matter (CDM) can play the role of the second component, on top of an early-time solution that raises~$H_0$.
In our previous works we studied the $\tilde\Lambda$CDM model, which is motivated by asymptotic safety~\cite{Weinberg:2009wa} and particle production~\cite{Parker:1973qd} in gravitational field theory and describes a vacuum energy interacting with matter and radiation~\cite{Xue:2014kna,Xue:2023qft}, and constrained it with scaling solutions of its rate equations~\cite{Gao:2021xnk,Gao:2022ahg,Zhang:2025dwu}.
Here we solve the rate equations directly, together with the linear perturbations, and restrict the exchange to the dark sector, while baryons and photons evolve as in \LCDM.
The coupling rate is set by a mass parameter~$\bar m$ and scales as a power~$H^{\alpha}$ of the Hubble rate with~$\alpha < 0$, so the exchange is negligible before recombination and becomes active only at low redshift.
The exchange therefore leaves the sound horizon and the early-time effect of the extra component unchanged.
The energy is transferred along the CDM four-velocity, so that the vacuum stays homogeneous in the CDM rest frame and CDM follows geodesics, as in the interacting vacuum--geodesic CDM scenario~\cite{Wands:2012vg,Wang:2014xca,Martinelli:2019dau,Hogg:2020rdp}.
When the vacuum decays into CDM, the newly created dark matter carries no density perturbation and dilutes the existing density contrast, so~$\Omega_m$ increases while~$\sigma_8$ decreases, and the net effect is a lower~$S_8$ (see ref.~\cite{Borges:2017jvi} for the growth of structure with geodesic CDM).
We add this exchange to the two early-time solutions discussed above.
With dark radiation, the vacuum exchanges energy with both CDM and the dark-radiation fluid, and we call the model \mbarLCDM; with EDE, the vacuum exchanges energy with CDM only, and we call the model \mbarEDE.

We constrain the models with the Planck 2018 TT,TE,EE+lowE+lensing CMB data~\cite{Planck:2018vyg,Planck:2019nip,Planck:2018lbu}, the DESI DR2 BAO measurements~\cite{DESI:2025zgx,DESI:2025zpo} together with 6dFGS~\cite{Beutler:2011hx} and SDSS MGS~\cite{Ross:2014qpa}, and the Pantheon+ SNe~Ia sample~\cite{Scolnic:2021amr,Brout:2022vxf}.
We refer to this combination as CBS, and to CBS plus the SH0ES Cepheid calibration~\cite{Riess:2021jrx} as CBSH.
To isolate the effect of the exchange, we compare \mbarLCDM{} with \LCDM{} plus the same dark-radiation fluid without exchange (\LCDM+DR), and \mbarEDE{} with EDE.
We leave the sign of the coupling free, so that the data determine the direction of the energy flow, and vary the power-law index~$\alpha$, which controls how late the exchange becomes active.
Fitting each model with and without the SH0ES calibration shows whether a preference for the exchange comes from the calibration or from the other data.
We find that in both models the data mildly favour an energy flow from the vacuum into CDM, more strongly without the SH0ES calibration, and that the exchange lowers~$S_8$ but does not raise~$H_0$.

A lower~$S_8$ does not necessarily imply a weaker cosmic-shear signal.
Since the exchange changes the matter density at low redshift, $\rho_m(z)$ no longer scales as~$(1+z)^3$, and~$S_8$, defined with the present-day~$\sigma_8$ and~$\Omega_m$, cannot be compared directly with cosmic-shear values inferred within \LCDM.
Cosmic shear is sensitive to the matter clustering over a range of redshifts set by the distances of the source galaxies, so what matters is how the exchange changes the clustering at those redshifts rather than today.
We therefore compute the effective lensing amplitude and the shear two-point correlation functions~$\xi_\pm$ with the matter density of each model, and compare them with the KiDS-Legacy measurements~\cite{Wright:2025xka}.
This comparison shows how much of the reduction in~$S_8$ translates into a weaker shear signal.

This paper is organised as follows.
Section~\ref{sec:model} describes the energy exchange, its effect on the background and on the growth of structure, and the two models built on it, \mbarLCDM{} and \mbarEDE.
Section~\ref{sec:data} presents the data, the models and their priors, and the sampling method.
In section~\ref{sec:results} we give the parameter constraints in the $H_0$--$S_8$ plane and their dependence on the power-law index~$\alpha$.
Section~\ref{sec:shear} computes the effective lensing amplitude and compares the predicted shear with KiDS-Legacy.
We discuss other approaches to the seesaw and the limitations of this work in section~\ref{sec:discussion}, and conclude in section~\ref{sec:conclusions}.
The appendices describe the numerical implementation, the validation of the cosmic-shear calculation and the convergence diagnostics.

\section{Energy exchange between the vacuum and dark matter}
\label{sec:model}

In this section we write the exchange for a dark sector made of CDM and, optionally, a dark-radiation fluid, and then add it to the two early-time solutions considered in this work.
We assume a spatially flat Friedmann--Lema\^itre--Robertson--Walker universe, use cosmic time~$t$ with~$\dot{} \equiv \mathrm{d}/\mathrm{d}t$ for the background, and conformal time~$\tau$ with~$' \equiv \mathrm{d}/\mathrm{d}\tau$ and~$\mathcal{H} \equiv a'/a = aH$ for the perturbations.

\subsection{Coupling and conservation equations}
\label{sec:model:coupling}

The vacuum energy, with energy density~$\rho_\Lambda$ and energy--momentum tensor~$T^{\mu\nu}_\Lambda = -\rho_\Lambda g^{\mu\nu}$, where~$g^{\mu\nu}$ is the inverse metric, exchanges energy with CDM and, in \mbarLCDM, with the dark radiation.
We denote by~$\rho_{\rm c}$ and~$\rho_{\rm dr}$ the energy densities of CDM and dark radiation, and by~$T^{\mu\nu}_{\rm c}$ and~$T^{\mu\nu}_{\rm dr}$ their energy--momentum tensors; in \mbarEDE{} there is no dark radiation and~$\rho_{\rm dr} = 0$.
Baryons, photons and neutrinos do not take part in the exchange and are separately conserved.
The conservation equations read
\begin{equation}
\nabla_\mu T^{\mu\nu}_\Lambda = Q^\nu,
\qquad
\nabla_\mu \left( T^{\mu\nu}_{\rm c} + T^{\mu\nu}_{\rm dr} \right) = -Q^\nu,
\label{eq:conservation}
\end{equation}
where~$\nabla_\mu$ is the covariant derivative and~$Q^\nu$ is the energy--momentum transfer four-vector, so that the total energy--momentum tensor is conserved.
We take the transfer to be parallel to the CDM four-velocity~$u^\nu_{\rm c}$, normalised as~$u_{{\rm c}\,\nu} u^\nu_{\rm c} = -1$,
\begin{equation}
Q^\nu = Q\, u^\nu_{\rm c},
\qquad
Q = \Gamma \left( \rho_{\rm c} + \rho_{\rm dr} \right),
\label{eq:Qnu}
\end{equation}
where~$Q$ is the energy transferred per unit volume and unit time, measured in the CDM rest frame, and~$\Gamma$ is the coupling rate, given in eq.~\eqref{eq:Gamma} below.
The dependence of~$Q$ on the densities follows the rate equations of the $\tilde\Lambda$CDM model~\cite{Xue:2023qft}, which describe the exchange between the vacuum and matter by analogy with a back-and-forth reaction such as~$e^+e^- \leftrightarrow \gamma\gamma$.
In these equations the energy that matter and radiation exchange with the vacuum is~$\Gamma$ times their energy density, as for a decay or an annihilation process, so that the exchange per unit volume scales with the amount of matter present.
We distribute the transfer between CDM and dark radiation in proportion to their energy densities.
The background equations are then
\begin{gather}
\dot\rho_\Lambda = \Gamma \left( \rho_{\rm c} + \rho_{\rm dr} \right),
\label{eq:rhoL}\\
\dot\rho_{\rm c} + 3H\rho_{\rm c} = -\Gamma \rho_{\rm c},
\label{eq:rhoc}\\
\dot\rho_{\rm dr} + 4H\rho_{\rm dr} = -\Gamma \rho_{\rm dr},
\label{eq:rhodr}
\end{gather}
so that CDM and dark radiation lose (or gain) the same fraction of their energy per unit time.

\begin{figure}[t]
\centering
\includegraphics[width=\textwidth]{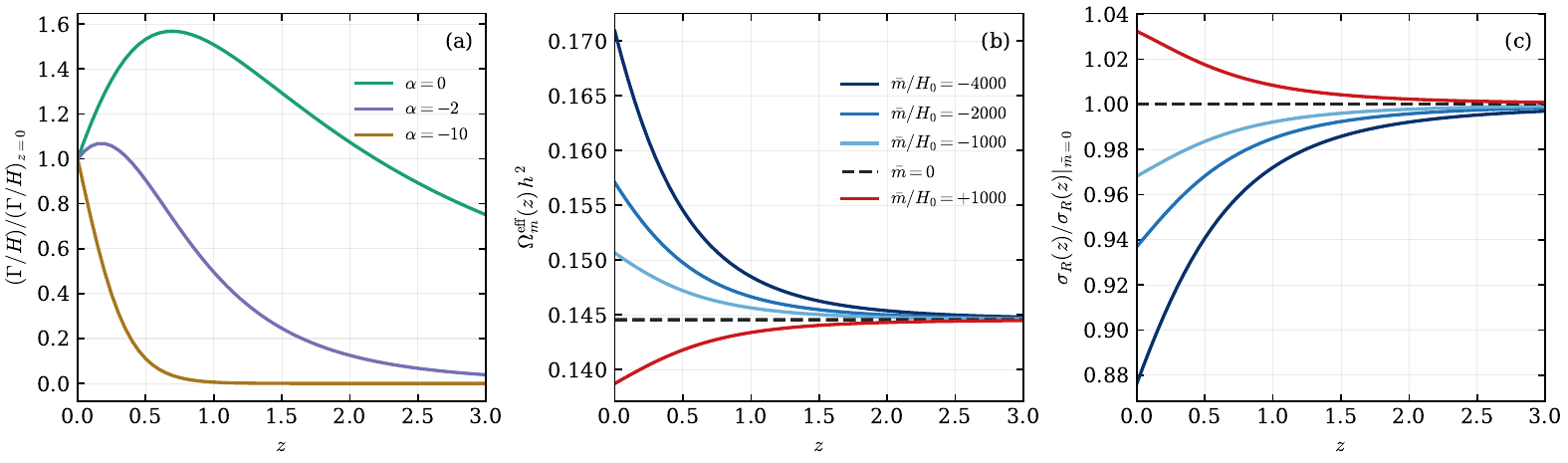}
\caption{Time dependence of the exchange and its effect on the background and on the growth of structure.
(a)~Coupling rate per Hubble time, eq.~\eqref{eq:GoH}, normalised to its present value, for~$\alpha = 0$, $-2$ and~$-10$; in the limit of weak coupling its shape does not depend on~$\bar m$.
(b) and~(c) vary~$\bar m$ for~$\alpha = -2$, with~$\omega_b$, the early-time CDM density~$\omega_{\rm cdm}F(a_*)$, the early-time~$\Delta N_{\rm eff} = 0.3$, $100\theta_s$, $A_s$, $n_s$ and~$\tau_{\rm reio}$ held fixed.
(b)~Physical matter density~$\Omega_m^{\rm eff}h^2$, eq.~\eqref{eq:Omeff}.
(c)~rms of the linear density field in spheres of fixed comoving radius~$R = 11.4$~Mpc ($8\,h^{-1}$Mpc for~$\bar m = 0$), divided by its value for~$\bar m = 0$; the fixed radius in Mpc removes the change of~$h$ with~$\bar m$~\cite{Sanchez:2020vvb}.
Negative~$\bar m$ (blue) corresponds to energy flowing from the vacuum into CDM.}
\label{fig:mbar_scan}
\end{figure}

The coupling rate is
\begin{equation}
\Gamma = \frac{\chi\,\epsilon}{4\pi}\, \bar m \left( \frac{H}{H_0} \right)^{\alpha},
\qquad
\epsilon \equiv -\frac{\dot H}{H^2} = \frac{3}{2}\,\frac{\rho_{\rm tot} + p_{\rm tot}}{\rho_{\rm tot}},
\label{eq:Gamma}
\end{equation}
where~$\rho_{\rm tot}$ and~$p_{\rm tot}$ include all components, and the vacuum contributes to~$\rho_{\rm tot}$ but not to~$\rho_{\rm tot} + p_{\rm tot}$, since~$p_\Lambda = -\rho_\Lambda$; $\bar m$ is a mass parameter that sets the strength and the direction of the exchange, $\alpha$ is a power-law index, and~$\chi \approx 1.85 \times 10^{-3}$ is the value obtained from the production of massive particle and antiparticle pairs~\cite{Xue:2021jyj,Xue:2020tpf}.
In the $\tilde\Lambda$CDM model the time variation of the vacuum energy follows from the scaling of the cosmological term in asymptotically safe gravity~\cite{Weinberg:2009wa,Xue:2014kna}, and the energy released by the vacuum is converted into massive particle--antiparticle pairs~\cite{Xue:2020tpf,Xue:2021jyj}; eq.~\eqref{eq:Gamma} parametrises this conversion, with~$\chi$ fixed by the pair-production rate, $\bar m$ treated as a free parameter, and the factor~$(H/H_0)^{\alpha}$ introduced phenomenologically to control when the exchange becomes active.
For~$\bar m > 0$ energy flows from the dark sector into the vacuum, and for~$\bar m < 0$ from the vacuum into the dark sector.
We quote~$\bar m$ in units of~$H_0$.
The rate per Hubble time,
\begin{equation}
\frac{\Gamma}{H} = \frac{\chi\,\epsilon}{4\pi}\, \frac{\bar m}{H_0} \left( \frac{H}{H_0} \right)^{\alpha - 1},
\label{eq:GoH}
\end{equation}
measures the fraction of the CDM energy transferred per $e$-fold.
Today~$\epsilon \simeq 0.5$, so~$\bar m/H_0 \simeq -2000$ gives~$\Gamma/H \simeq -0.15$.
At recombination~$H/H_0 \simeq 2 \times 10^4$, and for~$\alpha = -2$ the factor~$(H/H_0)^{\alpha-1}$ suppresses~$\Gamma/H$ to~$\sim 10^{-13}$.
For~$\alpha \le -2$ the exchange has no effect on the sound horizon, on recombination, or on the early-time effect of dark radiation or EDE.
For~$\alpha = 0$, $\Gamma/H$ is also small at early times, but in the radiation era~$\Gamma/H \propto a^2$, and the energy exchanged with the dark radiation changes~$\rho_\Lambda$ by an amount that grows as~$a^{-2}$ towards the past; we therefore switch the exchange on only at~$z < 10$.
At low redshift the index~$\alpha$ sets how fast the exchange switches on, as shown in panel~(a) of figure~\ref{fig:mbar_scan}.
At~$z = 0.5$, $\Gamma/H$ is about~$1.5$, $0.9$ and~$0.1$ times its present value for~$\alpha = 0$, $-2$ and~$-10$, respectively, and at~$z = 1$ about~$1.5$, $0.5$ and~$0.006$ times.

\subsection{Background evolution}
\label{sec:model:background}

Equations~\eqref{eq:rhoc} and~\eqref{eq:rhodr} integrate to
\begin{equation}
\rho_{\rm c}(a) = \rho_{\rm c,0}\, a^{-3} F(a),
\qquad
\rho_{\rm dr}(a) = \rho_{\rm dr,0}\, a^{-4} F(a),
\qquad
F(a) = \exp\!\left[ \int_a^1 \frac{\Gamma}{H}\, \mathrm{d}\ln a' \right],
\label{eq:F}
\end{equation}
where the subscript~0 denotes present-day values.
We use the present-day densities~$\omega_{\rm cdm}$ and~$N_{\rm dr}$ as input parameters and integrate the background backwards from today (appendix~\ref{app:class}).
For~$\bar m < 0$ the vacuum transfers energy to the dark sector, $F(a) < 1$ in the past, and~$\rho_{\rm c}a^3$ grows towards the present.
Since the CMB fixes the CDM density at early times, $\omega_{\rm cdm}F(a_*)$, the present-day~$\omega_{\rm cdm}$ and~$\Omega_m$ are larger than in a model without exchange.
For the same reason the contribution of dark radiation to the effective number of relativistic species before recombination is~$\Delta N_{\rm eff} = N_{\rm dr} F(a_*)$, smaller than~$N_{\rm dr}$.
The vacuum energy decreases with time, and the extra CDM changes the low-redshift expansion history, which is constrained by BAO and SNe~Ia.
At fixed angular acoustic scale the larger late-time matter density lowers~$H_0$: for~$\bar m/H_0 = -2000$ and~$\alpha = -2$, with the early-time parameters fixed as in figure~\ref{fig:mbar_scan}, $H_0$ decreases by~$1.4~\mathrm{km\,s^{-1}\,Mpc^{-1}}$.

The exchange also changes the redshift dependence of the matter density.
We define
\begin{equation}
\Omega_m^{\rm eff}(z) \equiv \frac{\rho_m(z)}{\rho_{\rm crit,0}\,(1+z)^3},
\label{eq:Omeff}
\end{equation}
where~$\rho_m$ includes baryons, CDM and massive neutrinos.
In \LCDM{} $\Omega_m^{\rm eff}$ is constant for~$z \lesssim 10$ apart from the small contribution of massive neutrinos, whereas for~$\bar m < 0$ it decreases with redshift.
The present-day~$\Omega_m$ therefore overestimates the matter density at the redshifts probed by cosmic shear, which we take into account in section~\ref{sec:shear}.

\subsection{Linear perturbations and the growth of structure}
\label{sec:model:perturbations}

Because~$Q^\nu$ is parallel to~$u^\nu_{\rm c}$, there is no momentum transfer in the CDM rest frame, and the CDM velocity obeys the same equation as in \LCDM.
In the synchronous gauge comoving with CDM, $\theta_{\rm c} = 0$ remains a solution, and CDM follows geodesics.
We further take the energy transfer to be unperturbed in this frame, $\delta Q = 0$, so that the vacuum stays homogeneous in the CDM rest frame and does not cluster.
This is the geodesic model of an interacting vacuum~\cite{Wands:2012vg,Wang:2014xca,Martinelli:2019dau,Hogg:2020rdp}.
Since the vacuum has no perturbations, the model avoids the large-scale instability that affects some interacting dark energy models and is otherwise treated with the parametrised post-Friedmann approach~\cite{Li:2014eha}.
The perturbed continuity equation for CDM becomes
\begin{equation}
\delta_{\rm c}' = -\frac{h'}{2} + a\Gamma\, \delta_{\rm c},
\label{eq:deltac}
\end{equation}
where~$h$ is the trace of the metric perturbation in the notation of ref.~\cite{Ma:1995ey}.
The last term follows from~$\delta\rho_{\rm c}$ being unchanged by the exchange while~$\bar\rho_{\rm c}$ changes at the rate~$\Gamma$.
For~$\bar m < 0$ the vacuum adds CDM uniformly in the CDM rest frame: the new dark matter carries no density perturbation and dilutes the existing density contrast~\cite{Borges:2017jvi}.
The metric perturbations are sourced by~$\delta\rho_{\rm c}$, which is not changed directly, so the growth of~$\delta_{\rm c}$ and of the baryon perturbations that follow it is slowed down.
With the primordial amplitude fixed by the CMB, $\sigma_8$ is lower and~$\Omega_m$ higher than without exchange.

Panels~(b) and~(c) of figure~\ref{fig:mbar_scan} illustrate these effects for~$\alpha = -2$, varying~$\bar m$ with the early-time parameters held fixed.
The exchange is active at~$z \lesssim 2$, where the physical matter density departs from its value without exchange and the amplitude of the density fluctuations at fixed comoving scale decreases.
For~$\bar m/H_0 = -2000$, $\sigma_8$ decreases from~$0.809$ to~$0.747$ and~$\Omega_m$ increases from~$0.292$ to~$0.330$.
The two changes act in opposite directions on~$S_8 = \sigma_8\sqrt{\Omega_m/0.3}$, and the decrease of~$\sigma_8$ dominates: $S_8$ decreases from~$0.797$ to~$0.784$.
The same behaviour appears in the posteriors of section~\ref{sec:results}.

\subsection{Combination with early-time solutions}
\label{sec:model:combination}

\subsubsection{Dark radiation: \texorpdfstring{\mbarLCDM}{mbarLCDM}}
\label{sec:model:dr}

In \mbarLCDM{} the vacuum exchanges energy with CDM and with a dark-radiation fluid, following eqs.~\eqref{eq:rhoL}--\eqref{eq:rhodr}.
We parametrise the present-day dark-radiation density as~$\rho_{\rm dr,0} = N_{\rm dr}\,\frac{7}{8}\left(\frac{4}{11}\right)^{4/3}\rho_{\gamma,0}$, so that~$N_{\rm dr}$ counts the dark radiation in units of one massless neutrino species.
The dark radiation raises the expansion rate before recombination and reduces the sound horizon.
We treat it as a perfect fluid with sound speed~$c_s^2 = 1/3$ and no anisotropic stress.
Since it does not free-stream, it does not produce the phase shift of the acoustic peaks caused by free-streaming neutrinos, and its effect on the CMB differs from that of extra neutrino species~\cite{Bashinsky:2003tk,Baumann:2015rya,Blinov:2020hmc}.
For~$\bar m = 0$ the model reduces to \LCDM+DR, which we use to isolate the effect of the exchange.

\subsubsection{Early dark energy: \texorpdfstring{\mbarEDE}{mbarEDE}}
\label{sec:model:ede}

In \mbarEDE{} the dark sector contains no dark radiation, and the vacuum exchanges energy with CDM only, i.e.\ eqs.~\eqref{eq:rhoL} and~\eqref{eq:rhoc} with~$\rho_{\rm dr} = 0$.
The EDE scalar field does not take part in the exchange.
It contributes to the energy density mainly around the critical redshift~$z_c \simeq 3700$, which we fix together with the initial field displacement (section~\ref{sec:data:models}), so that~$\Gamma/H$ is negligible when EDE is dynamically relevant and the two components act at separate epochs.
For~$\bar m = 0$ the model reduces to EDE.

\section{Data and methodology}
\label{sec:data}

\subsection{Data}
\label{sec:data:datasets}

For the CMB, we use the Planck 2018 likelihoods, Planck TT,TE,EE+lowE+lensing~\cite{Planck:2019nip,Planck:2018vyg}.
At high multipoles ($\ell \geq 30$), we use the full \texttt{Plik} likelihood for the TT, TE and EE power spectra, sampling all of its foreground and calibration nuisance parameters.
At low multipoles ($2 \leq \ell < 30$), we use the \texttt{Commander} likelihood for TT and the \texttt{SimAll} likelihood for EE.
For the lensing reconstruction, we use the baseline conservative multipole range~$8 \leq L \leq 400$~\cite{Planck:2018lbu}.

For BAO, we use the DESI DR2 measurements with their covariance matrix~\cite{DESI:2025zgx,DESI:2025zpo}, which comprise 13 data points at seven effective redshifts, $z = 0.295$, 0.51, 0.706, 0.934, 1.321, 1.484 and~2.33: one point at~$z = 0.295$ and two points at each of the other redshifts.
We also include one measurement each from 6dFGS at~$z = 0.106$~\cite{Beutler:2011hx} and from SDSS MGS at~$z = 0.15$~\cite{Ross:2014qpa}.
The MGS sample partly overlaps in redshift and sky area with the DESI BGS sample; we neglect the correlation between the two measurements, which has little effect because the MGS distance has an uncertainty of about~$4\%$, compared with about~$1\%$ for BGS.
In total, the BAO data consist of 15 data points.

For SNe~Ia, we use the Pantheon+ sample~\cite{Scolnic:2021amr,Brout:2022vxf}, which contains 1701 light curves.
Without SH0ES, we use the 1580 light curves at~$z > 0.01$ and treat the absolute magnitude~$M$ as a free parameter, so that the SNe~Ia constrain the shape of the expansion history but not~$H_0$.
With SH0ES, we use the joint Pantheon+ and SH0ES likelihood~\cite{Brout:2022vxf,Riess:2021jrx}, which includes the 1371 Hubble-flow SNe~Ia at~$z > 0.023$ and 77 light curves of SNe~Ia in Cepheid host galaxies, 1448 light curves in total; the Cepheid distances calibrate~$M$ and thus constrain~$H_0$.
Since the two SN samples differ, we do not compare~$\chi^2$ values between CBS and CBSH, but only differences between models within each data combination.
We do not impose a Gaussian prior on~$H_0$, because the SH0ES value of~$H_0$ is derived assuming a \LCDM-like expansion history at low redshift, which does not hold for models that modify the late-time expansion~\cite{Camarena:2021jlr,Efstathiou:2021ocp}, such as the models with energy exchange considered here.

\subsection{Models and priors}
\label{sec:data:models}

We compare \mbarLCDM{} with two models without energy exchange.
The first is \LCDM.
The second, \LCDM+DR, adds to \LCDM{} the same dark-radiation fluid as in \mbarLCDM, with abundance~$N_{\rm dr}$, but no exchange; the difference between \LCDM+DR and \mbarLCDM{} isolates the effect of the coupling.
In \mbarLCDM{} the coupling~$\bar m$ and~$N_{\rm dr}$ are free and the power-law index is fixed; we take~$\alpha = -2$ as the baseline, for which the exchange is confined to~$z \lesssim 2$ (figure~\ref{fig:mbar_scan}a), where it affects the growth of structure, and also consider~$\alpha = 0$ and~$-1$, for which it extends to higher redshift.
For~$\alpha = 0$ we switch the exchange on only at~$z < 10$, for the reason given in section~\ref{sec:model:coupling}.
At~$z = 10$ the posterior values of~$\bar m$ give~$|\Gamma/H| \sim 10^{-3}$, so the exchange is already weak where it is switched on.
Both signs of~$\bar m$ are allowed, so the data determine the direction of the energy flow.

\begin{table}[htbp]
\centering
\renewcommand{\arraystretch}{1.15}
\caption{Flat priors of the models with dark radiation.
``Unbounded'' means no prior boundary, and a dash means that the parameter is not varied.
$\omega_{\rm cdm}$ and~$N_{\rm dr}$ refer to present-day densities (section~\ref{sec:model:background}).
The Planck nuisance parameters and the SN~Ia absolute magnitude~$M$ are varied in all models.}
\label{tab:priors_dr}
\begin{tabular}{llll}
\hline
Parameter & \LCDM{} & \LCDM+DR & \mbarLCDM{} \\
\hline
$\omega_b$, $\omega_{\rm cdm}$, $100\theta_s$, $\ln(10^{10}A_s)$, $n_s$ & unbounded & unbounded & unbounded \\
$\tau_{\rm reio}$ & $> 0.004$ & $> 0.004$ & $> 0.004$ \\
$N_{\rm dr}$ & -- & $[0, 2]$ & $[0, 2]$ \\
$\bar m/H_0$ & -- & -- & $[-6000, 6000]$ \\
$\alpha$ (fixed) & -- & -- & $0$, $-1$, $-2$ \\
\hline
\end{tabular}
\end{table}

For EDE we take the axion-like potential~$V(\phi) = m^2 f^2 [1-\cos(\phi/f)]^3$~\cite{Kamionkowski:2014zda,Poulin:2018cxd,Smith:2019ihp}.
We vary only the maximum fractional contribution of EDE to the total energy density, $f_{\rm EDE}$, and fix the critical scale factor and the initial field displacement to~$\log_{10} a_c = -3.5691$ and~$\theta_i \equiv \phi_i/f = 2.775$, the best-fit values of the three-parameter model to the Planck power spectra, as in the one-parameter EDE of refs.~\cite{Murgia:2020ryi,Smith:2020rxx}.
For \mbarEDE{} (section~\ref{sec:model:ede}) we consider~$\alpha = -2$ and~$-10$.
Since more negative~$\alpha$ confines the exchange to lower redshift, a larger~$|\bar m|$ is needed for the same effect on~$\sigma_8$, and we widen the prior on~$\bar m$ accordingly for~$\alpha = -10$.

\begin{table}[htbp]
\centering
\renewcommand{\arraystretch}{1.15}
\caption{Flat priors of the EDE models.
Following the \texttt{AxiCLASS} setup, these models are sampled in~$H_0$ and~$A_s$ instead of~$100\theta_s$ and~$\ln(10^{10}A_s)$, with bounded priors on the base parameters.
The EDE parameters~$\log_{10}a_c = -3.5691$ and~$\theta_i = 2.775$ are fixed.
Other conventions as in table~\ref{tab:priors_dr}.}
\label{tab:priors_ede}
\begin{tabular}{llll}
\hline
Parameter & EDE & \mbarEDE{} ($\alpha = -2$) & \mbarEDE{} ($\alpha = -10$) \\
\hline
$\omega_b$ & $[0.018, 0.03]$ & $[0.018, 0.03]$ & $[0.018, 0.03]$ \\
$\omega_{\rm cdm}$ & $[0.1, 0.2]$ & $[0.1, 0.2]$ & $[0.1, 0.2]$ \\
$H_0~[\mathrm{km\,s^{-1}\,Mpc^{-1}}]$ & $[50, 80]$ & $[50, 80]$ & $[50, 80]$ \\
$10^9 A_s$ & $[1.8, 3]$ & $[1.8, 3]$ & $[1.8, 3]$ \\
$n_s$ & $[0.9, 1.1]$ & $[0.9, 1.1]$ & $[0.9, 1.1]$ \\
$\tau_{\rm reio}$ & $[0.004, 0.12]$ & $[0.004, 0.12]$ & $[0.004, 0.12]$ \\
$f_{\rm EDE}$ & $[10^{-4}, 0.3]$ & $[10^{-4}, 0.3]$ & $[10^{-4}, 0.3]$ \\
$\bar m/H_0$ & -- & $[-6000, 6000]$ & $[-2\times10^5, 2\times10^5]$ \\
\hline
\end{tabular}
\end{table}

We compute \LCDM, \LCDM+DR and \mbarLCDM{} with the Boltzmann code \texttt{CLASS}\footnote{\url{https://github.com/lesgourg/class_public}}~\cite{Lesgourgues:2011re,Blas:2011rf}, in which we implement the energy exchange (appendix~\ref{app:class}), and EDE and \mbarEDE{} with \texttt{AxiCLASS}\footnote{\url{https://github.com/PoulinV/AxiCLASS}}~\cite{Poulin:2018dzj,Smith:2019ihp}, to which we add the same exchange.
In all models we include one massive neutrino with~$m_\nu = 0.06$~eV, and compute the non-linear matter power spectrum with HMcode~\cite{Mead:2020vgs}, which enters the Planck lensing likelihood and the shear predictions of section~\ref{sec:shear}.
The free parameters of all models and their priors are listed in tables~\ref{tab:priors_dr} and~\ref{tab:priors_ede}.

\subsection{Sampling and model comparison}
\label{sec:data:sampling}

We sample the posterior distributions with \texttt{MontePython}~\cite{Audren:2012wb,Brinckmann:2018cvx}, using its Metropolis--Hastings algorithm.
For each model and data combination we run four chains, starting from proposal covariance matrices obtained in preliminary runs.
We stop the chains when the Gelman--Rubin statistic~\cite{Gelman:1992zz}, computed after removing the first~$30\%$ of each chain, satisfies~$R - 1 < 0.02$ for the cosmological parameters in two consecutive checks, and discard the same~$30\%$ as burn-in in the analysis.
With the burn-in removal and the estimator of \texttt{MontePython}, all runs have~$R - 1 < 0.025$, and twelve of the fifteen have~$R - 1 < 0.02$ (appendix~\ref{app:convergence}).

We compare the models through the minimum~$\chi^2 \equiv -2\ln\mathcal{L}$ found in the chains.
For each data combination we quote~$\Delta\chi^2$ relative to \LCDM{} and, for the models with energy exchange, relative to the corresponding model without exchange, \LCDM+DR for \mbarLCDM{} and EDE for \mbarEDE.
To account for the extra parameters we also quote the Akaike information criterion~\cite{Akaike:1974vps}, $\Delta{\rm AIC} = \Delta\chi^2 + 2\Delta k$, where~$\Delta k$ is the number of free parameters added to \LCDM: one for \LCDM+DR and EDE, and two for \mbarLCDM{} and \mbarEDE.
Since the minimum is estimated from the chains rather than with a dedicated minimiser, we do not interpret differences of order unity.

We quantify the agreement with external measurements of~$H_0$ and~$S_8$ in two ways.
For a single parameter we use~$(\mu - \mu_{\rm ext})/\sqrt{\sigma^2 + \sigma_{\rm ext}^2}$, where~$\mu$ and~$\sigma$ are the posterior mean and standard deviation and, for asymmetric external errors, $\sigma_{\rm ext}$ is the error on the side of~$\mu$.
For~$H_0$ and~$S_8$ jointly we use the parameter-shift statistic of ref.~\cite{Raveri:2021wfz}.
With~$\boldsymbol\theta = (H_0, S_8)$, the distribution of the shift~$\Delta\boldsymbol\theta = \boldsymbol\theta - \boldsymbol\theta_{\rm ext}$ is
\begin{equation}
P(\Delta\boldsymbol\theta) = \int \mathcal{P}(\boldsymbol\theta)\, \mathcal{P}_{\rm ext}(\boldsymbol\theta - \Delta\boldsymbol\theta)\, \mathrm{d}^2\theta ,
\label{eq:shift_pdf}
\end{equation}
where~$\mathcal{P}$ is our posterior and~$\mathcal{P}_{\rm ext}$ is the product of the two external likelihoods, treated as independent, each a Gaussian with the upper and lower errors on the two sides of the central value.
The probability that the shift lies in the region where~$P$ exceeds its value at zero shift,
\begin{equation}
\Delta_{\rm shift} = \int_{P(\Delta\boldsymbol\theta) > P(\boldsymbol 0)} P(\Delta\boldsymbol\theta)\, \mathrm{d}^2\Delta\theta ,
\label{eq:shift_prob}
\end{equation}
is converted into an equivalent number of Gaussian standard deviations, $n_\sigma = \sqrt{2}\,{\rm erf}^{-1}(\Delta_{\rm shift})$.
We evaluate eq.~\eqref{eq:shift_pdf} on a grid, from the weighted histogram of the chain samples.
As external values we take~$H_0 = 73.04 \pm 1.04~\mathrm{km\,s^{-1}\,Mpc^{-1}}$ from SH0ES~\cite{Riess:2021jrx} and~$S_8 = 0.815^{+0.016}_{-0.021}$ from KiDS-Legacy~\cite{Wright:2025xka}.
For CBSH, which includes the SH0ES calibration, these numbers measure the consistency of the combined posterior with the external values rather than the tension between independent data sets.
For the models with energy exchange the comparison in~$S_8$ is only indicative, since~$S_8$ is defined with present-day quantities (section~\ref{sec:model:background}); we compare with cosmic shear directly in section~\ref{sec:shear}.

\section{Results}
\label{sec:results}

\subsection{\texorpdfstring{The $H_0$--$S_8$ plane}{The H0-S8 plane}}
\label{sec:results:H0S8}

\begin{figure}[t]
\centering
\includegraphics[width=\textwidth]{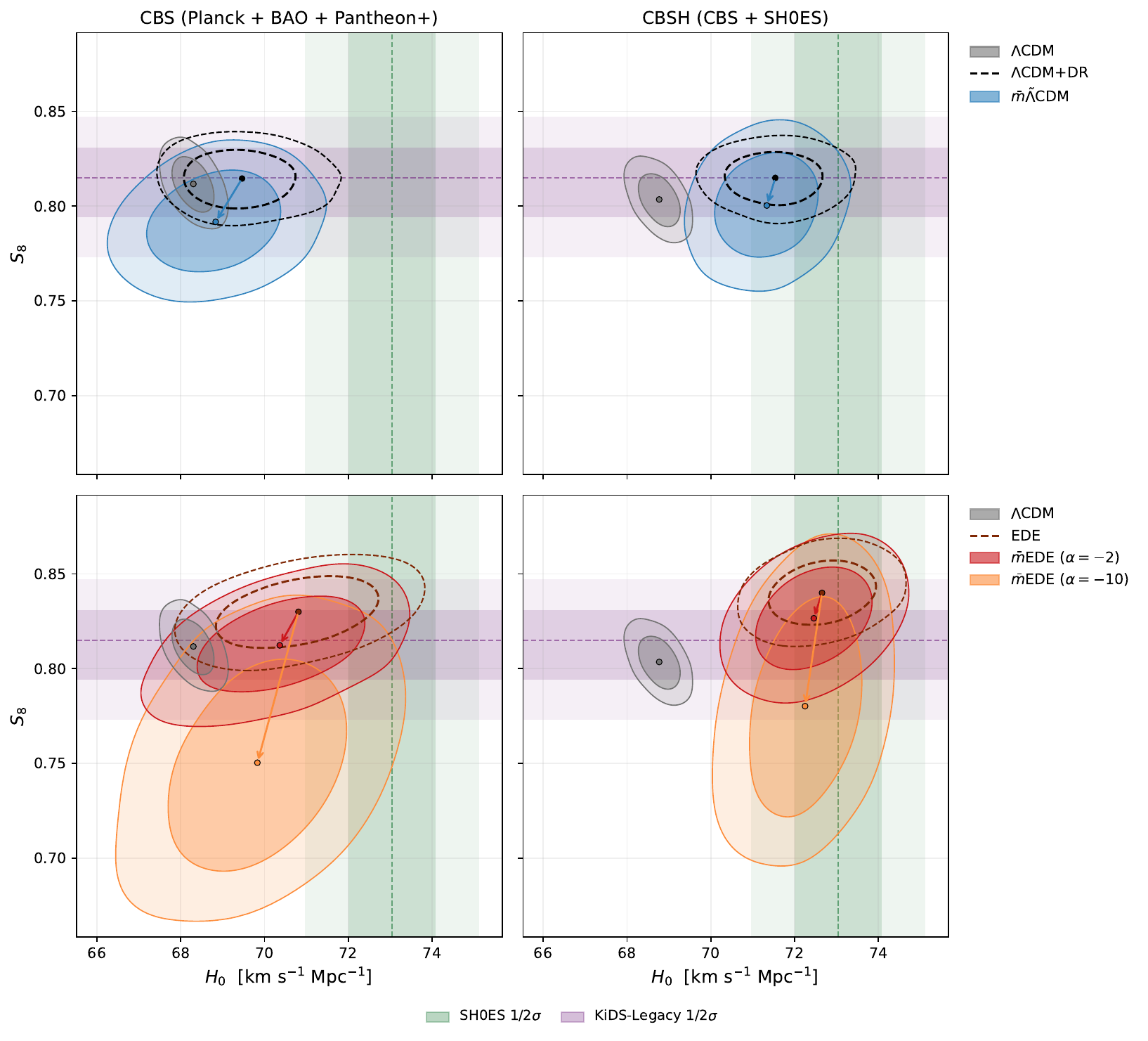}
\caption{Marginalised~$68\%$ and~$95\%$ contours in the $H_0$--$S_8$ plane for CBS (left) and CBSH (right).
Top: \LCDM, \LCDM+DR and \mbarLCDM{} ($\alpha = -2$); bottom: \LCDM, EDE and \mbarEDE{} ($\alpha = -2$ and~$-10$).
The points mark the posterior means, and the arrows connect each model without exchange to the corresponding model with exchange.
The vertical and horizontal bands show the~$1\sigma$ and~$2\sigma$ ranges of SH0ES~\cite{Riess:2021jrx} and KiDS-Legacy~\cite{Wright:2025xka}.}
\label{fig:H0S8}
\end{figure}

\begin{figure}[t]
\centering
\includegraphics[width=\textwidth]{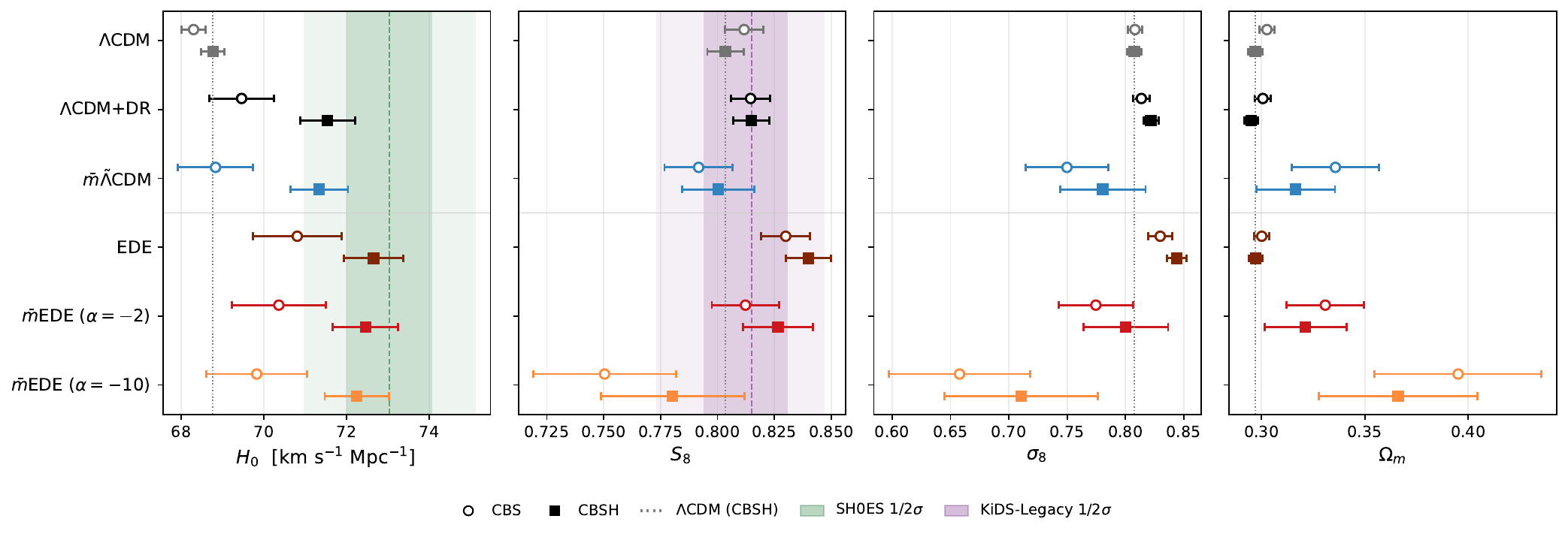}
\caption{Posterior means and~$68\%$ intervals of~$H_0$, $S_8$, $\sigma_8$ and~$\Omega_m$ for CBS (open circles) and CBSH (filled squares).
The dotted lines mark \LCDM{} for CBSH, and the bands show the~$1\sigma$ and~$2\sigma$ ranges of SH0ES and KiDS-Legacy.}
\label{fig:whisker}
\end{figure}

Figure~\ref{fig:H0S8} shows the marginalised posteriors in the $H_0$--$S_8$ plane, and figure~\ref{fig:whisker} summarises the constraints on~$H_0$, $S_8$, $\sigma_8$ and~$\Omega_m$.
Tables~\ref{tab:post_CBS} and~\ref{tab:post_CBSH} list the constraints on all cosmological parameters.
In \LCDM{} we find~$H_0 = 68.30 \pm 0.29~\mathrm{km\,s^{-1}\,Mpc^{-1}}$ and~$68.77 \pm 0.29~\mathrm{km\,s^{-1}\,Mpc^{-1}}$, and~$S_8 = 0.812 \pm 0.008$ and~$0.804 \pm 0.008$, for CBS and CBSH, respectively.

Both early-time solutions raise~$H_0$ and~$S_8$ together.
For CBS the dark radiation is not detected, $N_{\rm dr} < 0.42$ at~$95\%$ confidence level, and~$H_0$ increases to~$69.46 \pm 0.78~\mathrm{km\,s^{-1}\,Mpc^{-1}}$ with almost no change in~$S_8$.
For CBSH we find~$N_{\rm dr} = 0.52 \pm 0.11$ and~$H_0 = 71.59 \pm 0.68~\mathrm{km\,s^{-1}\,Mpc^{-1}}$; $\omega_{\rm cdm}$ increases from~$0.1171$ to~$0.1274$, and~$S_8$ from~$0.804$ to~$0.815$.
EDE raises~$H_0$ further, to~$70.81 \pm 1.08~\mathrm{km\,s^{-1}\,Mpc^{-1}}$ and~$72.65 \pm 0.71~\mathrm{km\,s^{-1}\,Mpc^{-1}}$ for CBS and CBSH, respectively, with~$f_{\rm EDE} = 0.086 \pm 0.033$ and~$0.138 \pm 0.020$.
For CBS and CBSH, respectively, the CDM density increases to~$\omega_{\rm cdm} = 0.127$ and~$0.134$, the spectral index to~$n_s = 0.983$ and~$0.992$, and~$S_8$ to~$0.830$ and~$0.840$, i.e.\ by~$0.018$ and~$0.036$ relative to \LCDM.

The exchange moves the posteriors to lower~$S_8$ at almost fixed~$H_0$, as indicated by the arrows in figure~\ref{fig:H0S8}.
For CBSH, \mbarLCDM{} gives~$S_8 = 0.800 \pm 0.016$, below both \LCDM+DR and \LCDM, while~$H_0 = 71.34 \pm 0.70~\mathrm{km\,s^{-1}\,Mpc^{-1}}$ decreases by only~$0.3~\mathrm{km\,s^{-1}\,Mpc^{-1}}$.
In \mbarEDE{} with~$\alpha = -2$ and~$-10$, $S_8$ decreases from~$0.840$ to~$0.827 \pm 0.015$ and~$0.780 \pm 0.032$, respectively, and~$H_0$ by~$0.2$ and~$0.4~\mathrm{km\,s^{-1}\,Mpc^{-1}}$.
For CBS the shifts in~$S_8$ are larger: $S_8 = 0.792 \pm 0.015$ for \mbarLCDM, and~$0.812 \pm 0.015$ and~$0.750 \pm 0.032$ for \mbarEDE{} with~$\alpha = -2$ and~$-10$, respectively.
Without the SH0ES calibration~$H_0$ is less tightly constrained and follows the coupling (section~\ref{sec:results:coupling}), and it decreases by~$0.6$, $0.4$ and~$1.0~\mathrm{km\,s^{-1}\,Mpc^{-1}}$, respectively.
In all models with exchange the posterior of~$S_8$ becomes broader, since~$S_8$ is correlated with the coupling, which the data constrain only weakly.

The lower~$S_8$ results from a lower~$\sigma_8$ and a higher~$\Omega_m$ (figure~\ref{fig:whisker}).
For CBSH, \mbarLCDM{} has~$\sigma_8 = 0.781 \pm 0.037$ and~$\Omega_m = 0.317 \pm 0.019$, compared with~$0.822 \pm 0.007$ and~$0.295 \pm 0.003$ in \LCDM+DR, and the decrease of~$\sigma_8$ dominates, as in the single-parameter scan of section~\ref{sec:model:perturbations}.
For \mbarEDE{} with~$\alpha = -10$ the changes are larger, $\sigma_8 = 0.711 \pm 0.066$ and~$\Omega_m = 0.366 \pm 0.038$.
These are present-day quantities, and we compare the models with cosmic shear directly in section~\ref{sec:shear}.

\begin{table}[t]
\centering
\footnotesize
\setlength{\tabcolsep}{4pt}
\renewcommand{\arraystretch}{1.15}
\caption{Posterior means and $68\%$ intervals of the parameters for CBS; upper limits are at the $95\%$ confidence level. $H_0$ is in units of~$\mathrm{km\,s^{-1}\,Mpc^{-1}}$; $\omega_{\rm cdm}$ and~$N_{\rm dr}$ are present-day values, and~$\Delta\chi^2_{\rm min}$ is relative to \LCDM.}
\label{tab:post_CBS}
\makebox[\textwidth][c]{%
\begin{tabular}{lcccccc}
\hline
Parameter & \LCDM & \LCDM+DR & \mbarLCDM & EDE & \mbarEDE & \mbarEDE \\
 &  &  & $\alpha=-2$ &  & $\alpha=-2$ & $\alpha=-10$ \\
\hline
$H_0$ & $68.30\pm0.29$ & $69.46\pm0.78$ & $68.83\pm0.91$ & $70.81\pm1.08$ & $70.36\pm1.13$ & $69.83\pm1.17$ \\
$S_8$ & $0.812\pm0.008$ & $0.815\pm0.009$ & $0.792\pm0.015$ & $0.830\pm0.011$ & $0.812\pm0.015$ & $0.750\pm0.032$ \\
$\sigma_8$ & $0.808\pm0.006$ & $0.814\pm0.007$ & $0.750\pm0.036$ & $0.830\pm0.010$ & $0.775\pm0.032$ & $0.656\pm0.061$ \\
$\Omega_m$ & $0.303\pm0.004$ & $0.301\pm0.004$ & $0.336\pm0.021$ & $0.300\pm0.004$ & $0.331\pm0.019$ & $0.396\pm0.040$ \\
$100\,\omega_b$ & $2.252\pm0.013$ & $2.267\pm0.015$ & $2.279\pm0.018$ & $2.266\pm0.014$ & $2.275\pm0.014$ & $2.269\pm0.014$ \\
$\omega_{\rm cdm}$ & $0.1180\pm0.0006$ & $0.1218\pm0.0024$ & $0.1355\pm0.0087$ & $0.1272\pm0.0039$ & $0.1403\pm0.0087$ & $0.1697\pm0.0175$ \\
$n_s$ & $0.9704\pm0.0034$ & $0.9714\pm0.0035$ & $0.9739\pm0.0036$ & $0.9826\pm0.0062$ & $0.9872\pm0.0069$ & $0.9840\pm0.0064$ \\
$\tau_{\rm reio}$ & $0.0596\pm0.0075$ & $0.0590\pm0.0077$ & $0.0617\pm0.0075$ & $0.0580\pm0.0069$ & $0.0603\pm0.0074$ & $0.0591\pm0.0073$ \\
$N_{\rm dr}$ & -- & $<0.42$ & $<0.59$ & -- & -- & -- \\
$f_{\rm EDE}$ & -- & -- & -- & $0.086\pm0.033$ & $0.101\pm0.034$ & $0.090\pm0.033$ \\
$\bar m/H_0$ & -- & -- & $-2103\pm1174$ & -- & $-1831\pm1020$ & $-21149\pm7809$ \\
\hline
$\Delta\chi^2_{\rm min}$ & $+0.0$ & $-0.9$ & $-4.6$ & $-6.0$ & $-7.8$ & $-13.2$ \\
\hline
\end{tabular}}
\end{table}

\begin{table}[t]
\centering
\footnotesize
\setlength{\tabcolsep}{4pt}
\renewcommand{\arraystretch}{1.15}
\caption{Posterior means and $68\%$ intervals of the parameters for CBSH; upper limits are at the $95\%$ confidence level. $H_0$ is in units of~$\mathrm{km\,s^{-1}\,Mpc^{-1}}$; $\omega_{\rm cdm}$ and~$N_{\rm dr}$ are present-day values, and~$\Delta\chi^2_{\rm min}$ is relative to \LCDM.}
\label{tab:post_CBSH}
\makebox[\textwidth][c]{%
\begin{tabular}{lcccccc}
\hline
Parameter & \LCDM & \LCDM+DR & \mbarLCDM & EDE & \mbarEDE & \mbarEDE \\
 &  &  & $\alpha=-2$ &  & $\alpha=-2$ & $\alpha=-10$ \\
\hline
$H_0$ & $68.77\pm0.29$ & $71.59\pm0.68$ & $71.34\pm0.70$ & $72.65\pm0.71$ & $72.46\pm0.79$ & $72.25\pm0.78$ \\
$S_8$ & $0.804\pm0.008$ & $0.815\pm0.008$ & $0.800\pm0.016$ & $0.840\pm0.010$ & $0.827\pm0.015$ & $0.780\pm0.032$ \\
$\sigma_8$ & $0.808\pm0.006$ & $0.822\pm0.007$ & $0.781\pm0.037$ & $0.844\pm0.008$ & $0.800\pm0.036$ & $0.711\pm0.066$ \\
$\Omega_m$ & $0.297\pm0.004$ & $0.295\pm0.003$ & $0.317\pm0.019$ & $0.297\pm0.003$ & $0.321\pm0.020$ & $0.366\pm0.038$ \\
$100\,\omega_b$ & $2.264\pm0.013$ & $2.297\pm0.014$ & $2.307\pm0.018$ & $2.276\pm0.012$ & $2.285\pm0.013$ & $2.281\pm0.012$ \\
$\omega_{\rm cdm}$ & $0.1171\pm0.0006$ & $0.1274\pm0.0023$ & $0.1373\pm0.0091$ & $0.1335\pm0.0029$ & $0.1451\pm0.0098$ & $0.1675\pm0.0189$ \\
$n_s$ & $0.9727\pm0.0034$ & $0.9737\pm0.0033$ & $0.9759\pm0.0038$ & $0.9921\pm0.0045$ & $0.9963\pm0.0058$ & $0.9947\pm0.0048$ \\
$\tau_{\rm reio}$ & $0.0627\pm0.0075$ & $0.0596\pm0.0069$ & $0.0619\pm0.0082$ & $0.0572\pm0.0068$ & $0.0597\pm0.0075$ & $0.0584\pm0.0067$ \\
$N_{\rm dr}$ & -- & $0.52\pm0.11$ & $0.64\pm0.17$ & -- & -- & -- \\
$f_{\rm EDE}$ & -- & -- & -- & $0.138\pm0.020$ & $0.154\pm0.024$ & $0.150\pm0.022$ \\
$\bar m/H_0$ & -- & -- & $-1367\pm1190$ & -- & $-1473\pm1158$ & $-15876\pm8074$ \\
\hline
$\Delta\chi^2_{\rm min}$ & $+0.0$ & $-18.8$ & $-21.8$ & $-33.4$ & $-34.1$ & $-36.8$ \\
\hline
\end{tabular}}
\end{table}

\subsection{Constraints on the coupling}
\label{sec:results:coupling}

\begin{figure}[t]
\centering
\includegraphics[width=\textwidth]{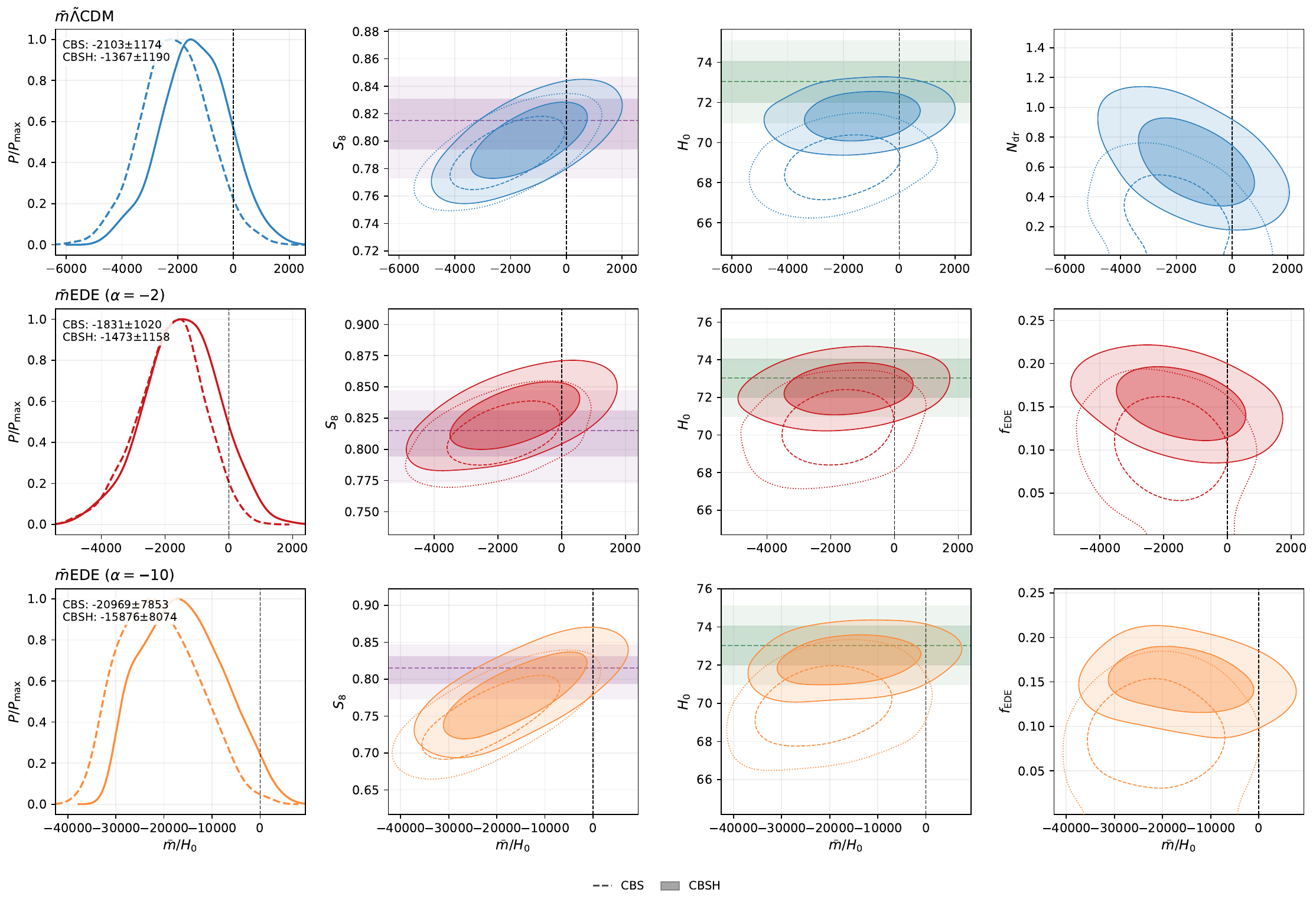}
\caption{Posterior of~$\bar m/H_0$ (first column) and its correlations with~$S_8$, $H_0$ and~$N_{\rm dr}$ or~$f_{\rm EDE}$, for \mbarLCDM{} (top) and \mbarEDE{} with~$\alpha = -2$ (middle) and~$\alpha = -10$ (bottom), for CBS (dashed) and CBSH (solid lines and filled contours).
The vertical dotted lines mark~$\bar m = 0$, and the bands show the~$1\sigma$ and~$2\sigma$ ranges of SH0ES and KiDS-Legacy.}
\label{fig:mbar}
\end{figure}

Figure~\ref{fig:mbar} shows the posterior of~$\bar m/H_0$ and its correlations with~$S_8$, $H_0$ and the parameter of the early-time solution.
In all six cases the posterior peaks at~$\bar m < 0$, i.e.\ at an energy flow from the vacuum into the dark sector.
We quote the significance of a non-zero coupling as the posterior mean divided by the standard deviation, together with the posterior probability of~$\bar m < 0$.
For the skewed posteriors of~$\alpha = -10$ the ratio is only indicative and~$P(\bar m < 0)$ is the more direct measure; both depend on the prior range of~$\bar m$ (section~\ref{sec:discussion:caveats}).
For CBS we find~$\bar m/H_0 = -2100 \pm 1200$ for \mbarLCDM, $1.8\sigma$ from zero with~$P(\bar m < 0) = 0.96$, and~$-1830 \pm 1020$ ($1.8\sigma$, $P = 0.97$) and~$-21000 \pm 7800$ ($2.7\sigma$, $P = 0.994$) for \mbarEDE{} with~$\alpha = -2$ and~$-10$, respectively.
Adding SH0ES weakens the preference: for CBSH we find~$-1370 \pm 1190$ ($1.1\sigma$, $P = 0.87$), $-1470 \pm 1160$ ($1.3\sigma$, $P = 0.90$) and~$-15900 \pm 8100$ ($2.0\sigma$, $P = 0.975$) for \mbarLCDM{} and \mbarEDE{} with~$\alpha = -2$ and~$-10$, respectively.
For~$\alpha = -2$ the preferred values correspond to a present rate~$\Gamma/H \simeq -0.1$ to~$-0.15$ (figure~\ref{fig:late}), so that the vacuum adds about~$10$--$15\%$ of the CDM energy per $e$-fold today.
For~$\alpha = -10$ the present rate is larger, $\Gamma/H \simeq -1.9$ (CBS) and~$-1.2$ (CBSH), but it falls below~$10^{-2}$ in absolute value at~$z \simeq 1$, and the exchange is confined to~$z \lesssim 0.7$.

\begin{figure}[t]
\centering
\includegraphics[width=0.85\textwidth]{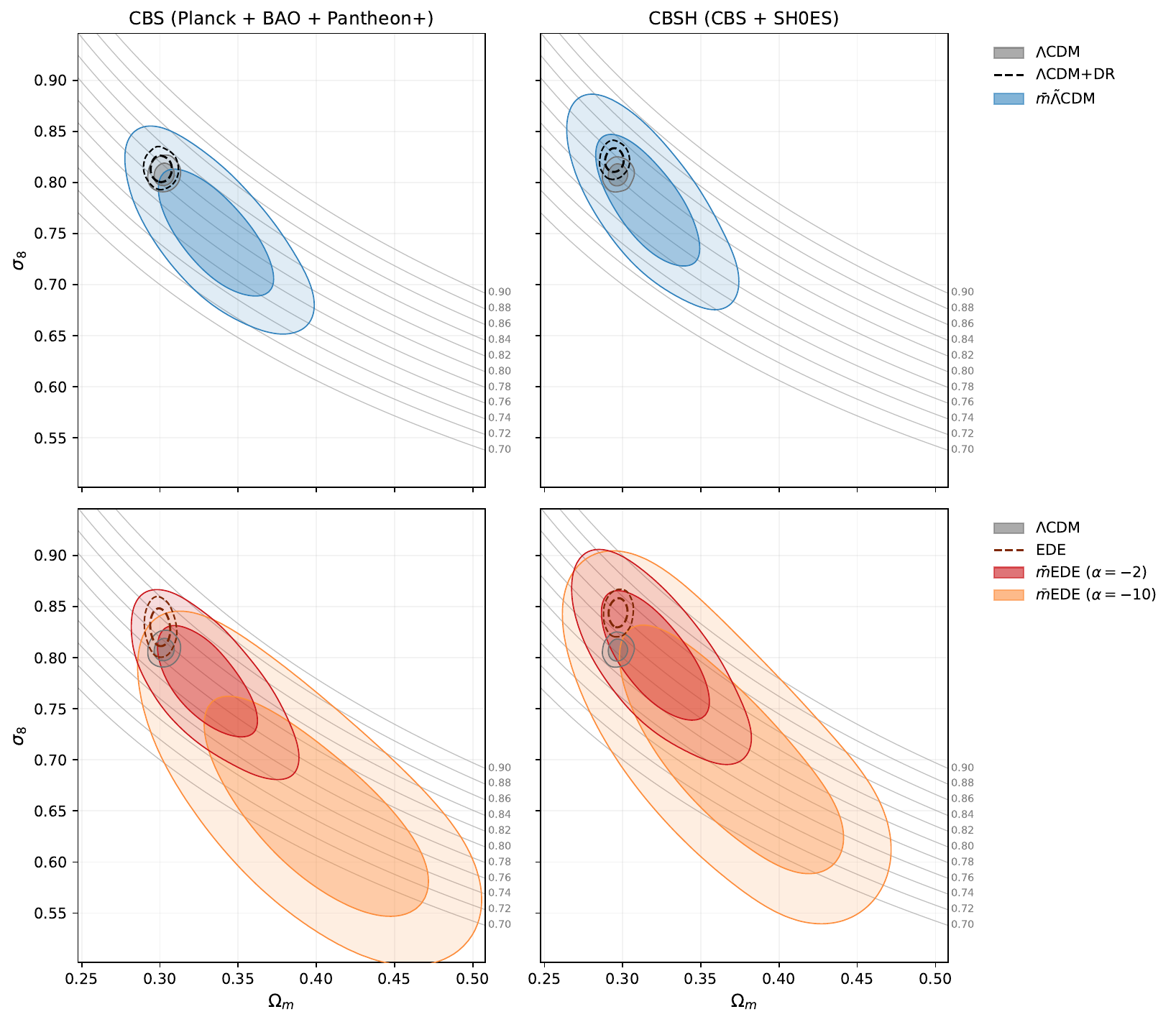}
\caption{Marginalised~$68\%$ and~$95\%$ contours in the $\sigma_8$--$\Omega_m$ plane, with the same layout as figure~\ref{fig:H0S8}.
The grey lines are lines of constant~$S_8$.}
\label{fig:s8Om}
\end{figure}

The coupling is strongly correlated with~$\sigma_8$ and anti-correlated with~$\Omega_m$: the correlation coefficients are~$0.94$--$0.99$ and~$-0.98$ to~$-0.99$ in all six cases.
The correlation with~$S_8$ is weaker, $0.64$--$0.95$, because the changes in~$\sigma_8$ and~$\Omega_m$ partly compensate, and the correlation with~$H_0$ is only~$0.2$--$0.4$.
In the $\sigma_8$--$\Omega_m$ plane (figure~\ref{fig:s8Om}) the posteriors of the models with exchange extend from the region of the models without exchange towards lower~$\sigma_8$ and higher~$\Omega_m$, and cross the lines of constant~$S_8$ towards lower values.
For \mbarLCDM{} the coupling is anti-correlated with~$N_{\rm dr}$, and for CBSH the posterior mean of~$N_{\rm dr}$ increases from~$0.52$ to~$0.64$.
Since~$N_{\rm dr}$ is defined today and the vacuum also transfers energy to the dark radiation, part of this increase is produced at low redshift and does not contribute to the early expansion (section~\ref{sec:model:background}).
In \mbarEDE{} $f_{\rm EDE}$ changes little, from~$0.138$ to~$0.154$ and~$0.150$ for CBSH.

\subsection{Model comparison}
\label{sec:results:comparison}

Table~\ref{tab:comparison} lists~$\Delta\chi^2$ and~$\Delta{\rm AIC}$ and the agreement with the external measurements of~$H_0$ and~$S_8$.
For CBS, all extensions improve the fit only moderately.
The exchange lowers~$\chi^2$ relative to the corresponding model without exchange by~$3.7$ for \mbarLCDM{} and by~$1.9$ and~$7.2$ for \mbarEDE{} with~$\alpha = -2$ and~$-10$.
\mbarEDE{} with~$\alpha = -10$ gives the largest improvement over \LCDM, $\Delta\chi^2 = -13.2$ and~$\Delta{\rm AIC} = -9.2$, followed by EDE and \mbarEDE{} with~$\alpha = -2$, $\Delta{\rm AIC} \simeq -4$; for \LCDM+DR and \mbarLCDM{} $\Delta{\rm AIC}$ is close to zero.
For CBSH the extensions that raise~$H_0$ improve the fit substantially, by~$\Delta\chi^2 = -18.8$ for \LCDM+DR and~$-33.4$ for EDE, and the exchange adds~$-3.0$, $-0.7$ and~$-3.3$ for \mbarLCDM{} and \mbarEDE{} with~$\alpha = -2$ and~$-10$, respectively.
The lowest~$\chi^2$ is obtained for \mbarEDE{} with~$\alpha = -10$, $\Delta\chi^2 = -36.8$ and~$\Delta{\rm AIC} = -32.8$, but its difference from EDE in~$\Delta{\rm AIC}$, $\Delta\chi^2_{\rm ex} + 2 = -1.3$ before rounding, is within the uncertainty of the minimum estimated from the chains.

For CBS, \LCDM{} is in~$4.4\sigma$ tension with SH0ES and in~$4.0\sigma$ joint tension with SH0ES and KiDS-Legacy.
EDE reduces the joint tension to~$1.4\sigma$, and \mbarEDE{} with~$\alpha = -2$ to~$1.3\sigma$, the lowest value for CBS.
In the dark-radiation models the joint tension remains at~$2.2$--$2.6\sigma$, since~$H_0$ stays below~$70~\mathrm{km\,s^{-1}\,Mpc^{-1}}$.
For CBSH, the joint tension decreases from~$3.6\sigma$ in \LCDM{} to below~$1\sigma$ in all extensions, and is lowest for \mbarEDE: $0.36\sigma$ for~$\alpha = -2$ and~$0.49\sigma$ for~$\alpha = -10$, compared with~$0.91\sigma$ for EDE.
The improvement comes from~$S_8$: EDE is~$1.3\sigma$ above KiDS-Legacy, and the exchange reduces this to~$0.5\sigma$ for~$\alpha = -2$, while the agreement in~$H_0$ changes little.
Since the KiDS-Legacy value of~$S_8$ is close to that of \LCDM, a lower~$S_8$ does not always improve the agreement: for \mbarEDE{} with~$\alpha = -10$ and CBS, $S_8$ lies~$1.7\sigma$ below KiDS-Legacy.
Compared with the lower KiDS-1000 value, $S_8 = 0.759^{+0.024}_{-0.021}$~\cite{KiDS:2020suj}, the exchange reduces the difference in all models, for example for CBSH from~$3.1\sigma$ to~$2.4\sigma$ and~$0.5\sigma$ in the EDE models with~$\alpha = -2$ and~$-10$.

\begin{table}[t]
\centering
\renewcommand{\arraystretch}{1.15}
\caption{Model comparison.
$\Delta\chi^2$ and~$\Delta{\rm AIC}$ are relative to \LCDM{} for the same data, and~$\Delta\chi^2_{\rm ex}$ is relative to the corresponding model without exchange (\LCDM+DR for \mbarLCDM, EDE for \mbarEDE).
The significance of the coupling is the posterior mean of~$\bar m$ divided by its standard deviation.
The last three columns give the agreement with SH0ES in~$H_0$, with KiDS-Legacy in~$S_8$ (positive when the model is higher), and the joint tension in~$(H_0, S_8)$ (section~\ref{sec:data:sampling}).
For CBSH, the~$H_0$ and joint columns measure the consistency of a posterior that includes SH0ES.}
\label{tab:comparison}
\begin{tabular}{lccccccc}
\hline
Model & $\Delta\chi^2$ & $\Delta{\rm AIC}$ & $\Delta\chi^2_{\rm ex}$ & $\bar m \neq 0$ & $H_0$ & $S_8$ & Joint \\
\hline
\multicolumn{8}{c}{CBS} \\
\hline
\LCDM & $0$ & $0$ & -- & -- & $-4.4\sigma$ & $-0.1\sigma$ & $4.0\sigma$ \\
\LCDM+DR & $-0.9$ & $+1.1$ & -- & -- & $-2.8\sigma$ & $-0.0\sigma$ & $2.2\sigma$ \\
\mbarLCDM & $-4.6$ & $-0.6$ & $-3.7$ & $1.8\sigma$ & $-3.1\sigma$ & $-0.9\sigma$ & $2.6\sigma$ \\
EDE & $-6.0$ & $-4.0$ & -- & -- & $-1.5\sigma$ & $+0.8\sigma$ & $1.4\sigma$ \\
\mbarEDE{} ($\alpha = -2$) & $-7.8$ & $-3.8$ & $-1.9$ & $1.8\sigma$ & $-1.7\sigma$ & $-0.1\sigma$ & $1.3\sigma$ \\
\mbarEDE{} ($\alpha = -10$) & $-13.2$ & $-9.2$ & $-7.2$ & $2.7\sigma$ & $-2.0\sigma$ & $-1.7\sigma$ & $1.8\sigma$ \\
\hline
\multicolumn{8}{c}{CBSH} \\
\hline
\LCDM & $0$ & $0$ & -- & -- & $-4.0\sigma$ & $-0.5\sigma$ & $3.6\sigma$ \\
\LCDM+DR & $-18.8$ & $-16.8$ & -- & -- & $-1.2\sigma$ & $-0.0\sigma$ & $0.67\sigma$ \\
\mbarLCDM & $-21.8$ & $-17.8$ & $-3.0$ & $1.1\sigma$ & $-1.4\sigma$ & $-0.6\sigma$ & $0.90\sigma$ \\
EDE & $-33.4$ & $-31.4$ & -- & -- & $-0.3\sigma$ & $+1.3\sigma$ & $0.91\sigma$ \\
\mbarEDE{} ($\alpha = -2$) & $-34.1$ & $-30.1$ & $-0.7$ & $1.3\sigma$ & $-0.4\sigma$ & $+0.5\sigma$ & $0.36\sigma$ \\
\mbarEDE{} ($\alpha = -10$) & $-36.8$ & $-32.8$ & $-3.3$ & $2.0\sigma$ & $-0.6\sigma$ & $-0.9\sigma$ & $0.49\sigma$ \\
\hline
\end{tabular}
\end{table}

\subsection{\texorpdfstring{Dependence on $\alpha$ and on the direction of the energy flow}{Dependence on alpha and on the direction of the energy flow}}
\label{sec:results:alpha}

\begin{figure}[t]
\centering
\includegraphics[width=\textwidth]{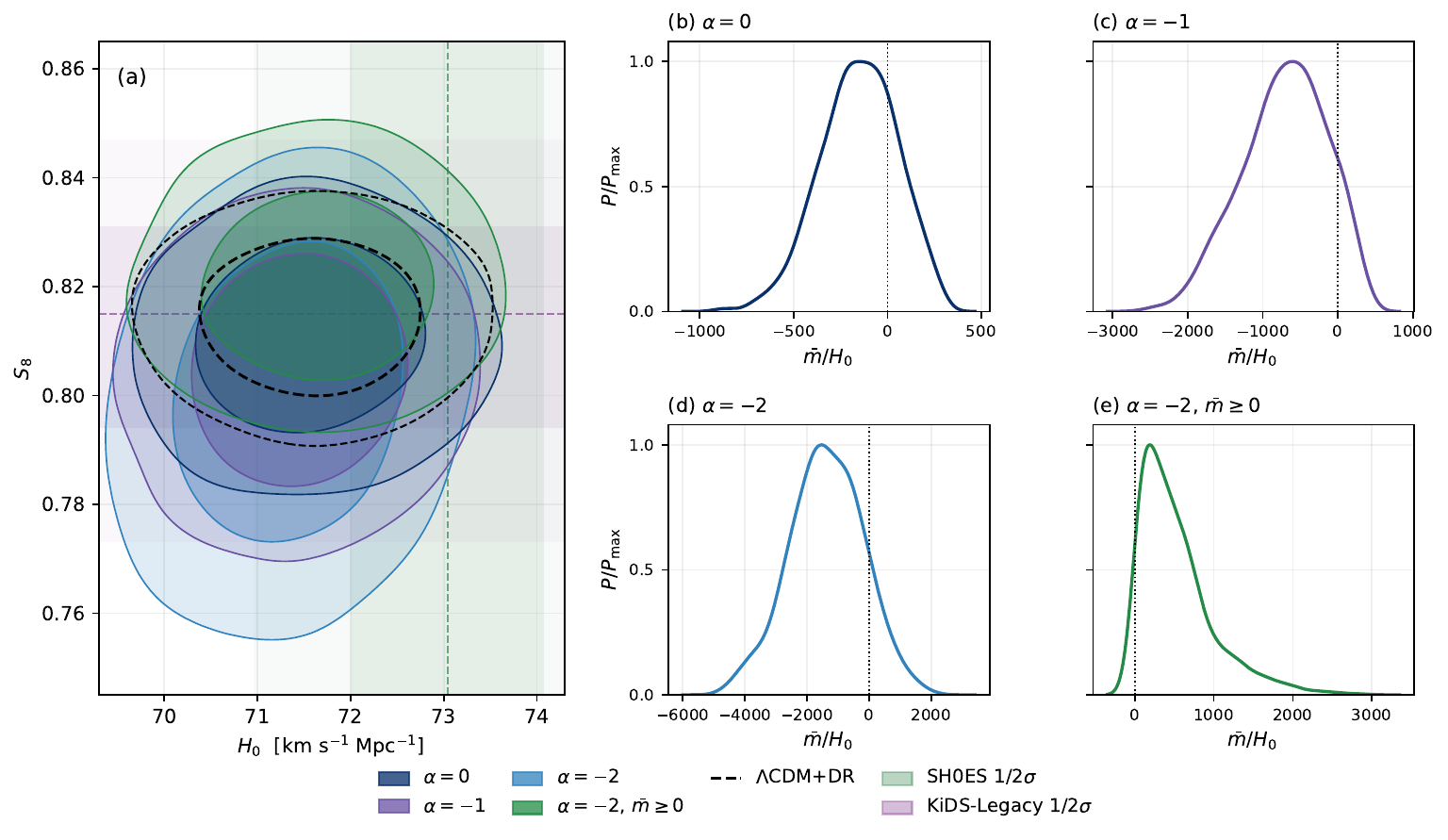}
\caption{\mbarLCDM{} for CBSH with different power-law indices and directions of the energy flow.
(a)~$68\%$ (dark) and~$95\%$ (light) contours in the $H_0$--$S_8$ plane for~$\alpha = 0$, $-1$, $-2$, and~$\alpha = -2$ with~$\bar m \ge 0$, compared with \LCDM+DR (dashed).
(b)--(e)~The corresponding posteriors of~$\bar m/H_0$.
The bands show the~$1\sigma$ and~$2\sigma$ ranges of SH0ES and KiDS-Legacy.}
\label{fig:alpha}
\end{figure}

Figure~\ref{fig:alpha} shows \mbarLCDM{} for CBSH with~$\alpha = 0$, $-1$ and~$-2$, and with~$\alpha = -2$ and the prior restricted to~$\bar m \ge 0$.
The constraint on~$\bar m$ tightens as~$\alpha$ increases: $\bar m/H_0 = -150 \pm 210$ for~$\alpha = 0$, $-710 \pm 560$ for~$\alpha = -1$, and~$-1370 \pm 1190$ for~$\alpha = -2$.
For larger~$\alpha$ the exchange is active over a wider range of redshift (figure~\ref{fig:mbar_scan}a), where BAO and SNe~Ia constrain the expansion history, and a smaller coupling is allowed.
The reduction of~$S_8$ relative to \LCDM+DR therefore increases with decreasing~$\alpha$, from~$0.004$ for~$\alpha = 0$ to~$0.010$ and~$0.014$ for~$\alpha = -1$ and~$-2$, respectively, and the improvement in~$\chi^2$ from~$1.3$ to~$1.8$ and~$3.0$.
The same trend holds for \mbarEDE: with~$\alpha = -10$ the exchange is confined to~$z \lesssim 0.7$, and both the reduction of~$S_8$ and the improvement in~$\chi^2$ are larger than for~$\alpha = -2$ (section~\ref{sec:results:H0S8}).

If the energy is allowed to flow only from the dark sector into the vacuum, $\bar m \ge 0$, the posterior peaks close to zero, $\bar m/H_0 = 530 \pm 470$.
The vacuum then removes CDM at late times, so that~$\Omega_m$ decreases to~$0.287 \pm 0.007$ and~$S_8$ increases to~$0.821 \pm 0.010$, above \LCDM+DR.
The fit is not improved relative to \LCDM+DR, $\Delta\chi^2 = +0.4$.
An energy flow from the vacuum into the dark sector lowers~$S_8$ and improves the fit, whereas the opposite flow raises~$S_8$ and does not.

\subsection{\texorpdfstring{Growth rate $f\sigma_8(z)$}{Growth rate f sigma8(z)}}
\label{sec:results:fs8}

\begin{figure}[t]
\centering
\includegraphics[width=\textwidth]{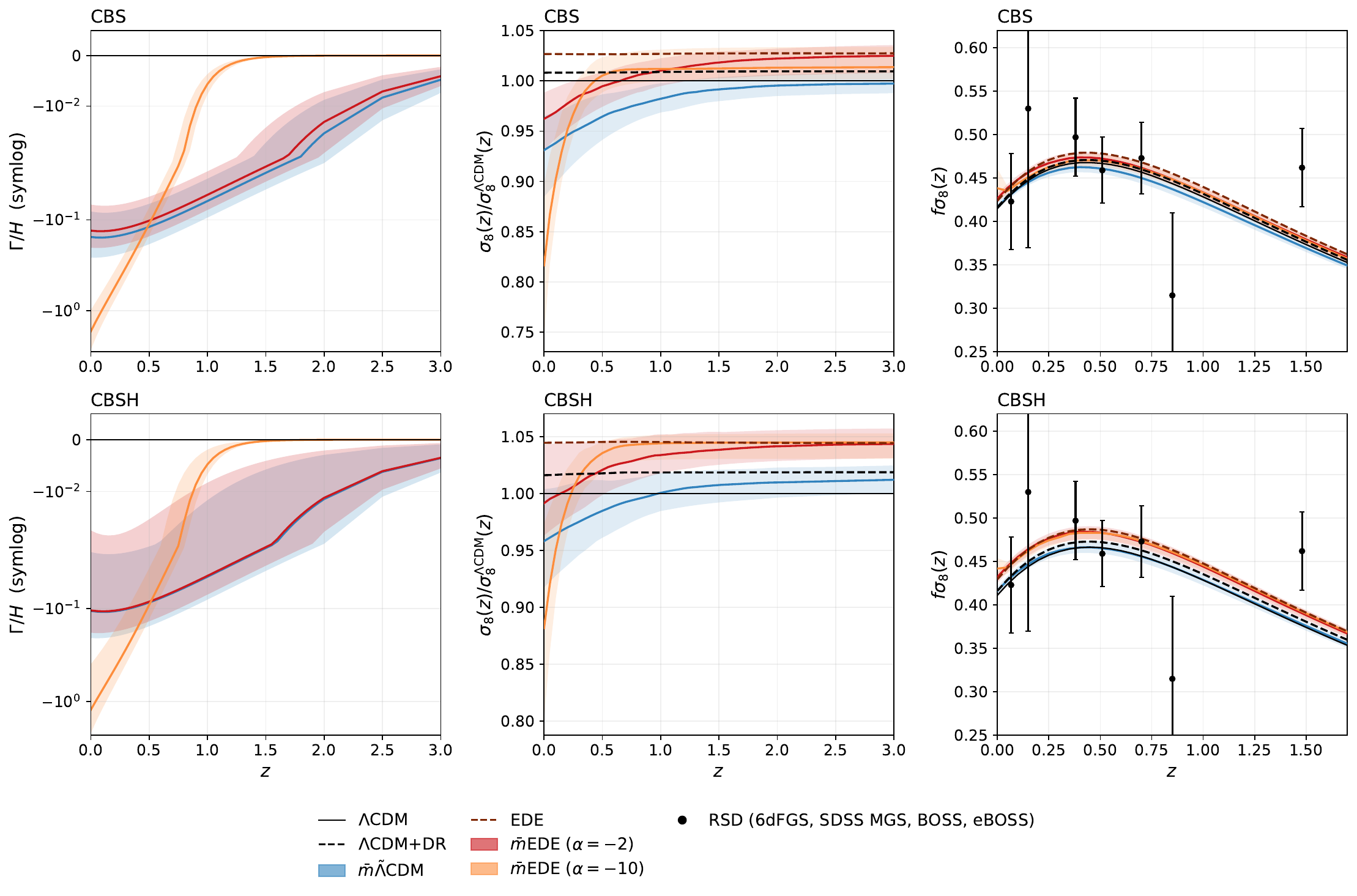}
\caption{Late-time evolution for CBS (top) and CBSH (bottom).
Left: coupling rate per Hubble time~$\Gamma/H$ for the models with exchange, on a symmetric logarithmic scale.
Middle: $\sigma_8(z)$ relative to the median of \LCDM.
Right: $f\sigma_8(z)$, with~$f = f_v$ of eq.~\eqref{eq:fv} for the models with exchange, compared with the RSD measurements from 6dFGS, SDSS MGS, BOSS DR12 and eBOSS DR16~\cite{Beutler:2012px,Howlett:2014opa,eBOSS:2020yzd}.
Lines and bands show the median and the~$68\%$ range of~$80$ posterior samples; for \LCDM{} and the models without exchange only the median is shown.}
\label{fig:late}
\end{figure}

Figure~\ref{fig:late} shows the redshift dependence of the coupling rate, of~$\sigma_8(z)$ relative to \LCDM, and of~$f\sigma_8(z)$, computed for~$80$ samples of each posterior.
The reduction of~$\sigma_8(z)$ is concentrated at low redshift.
For CBSH, $\sigma_8(z)$ in \mbarLCDM{} is~$4\%$ below \LCDM{} at~$z = 0$, $1.5\%$ below at~$z = 0.5$ and equal to it at~$z = 1$.
In \mbarEDE{} with~$\alpha = -10$, $\sigma_8(z)$ is~$12\%$ below \LCDM{} at~$z = 0$, but at~$z \ge 0.5$ it is~$4\%$ above, close to EDE without exchange.
Probes of the clustering amplitude at~$z \gtrsim 0.5$ are much less affected by the exchange than the present-day~$\sigma_8$.

Redshift-space distortions measure the peculiar velocities, which are related to the growth of the density contrast through the continuity equation.
With the exchange, the matter density contrast obeys~$\delta_m' = -\theta_m + a\Gamma(\rho_{\rm c}/\rho_m)\delta_m$ on sub-horizon scales, from eq.~\eqref{eq:deltac}, and the velocity growth rate differs from the density growth rate~$f \equiv \mathrm{d}\ln\delta_m/\mathrm{d}\ln a$:
\begin{equation}
f_v \equiv -\frac{\theta_m}{aH\delta_m} = f - \frac{\rho_{\rm c}}{\rho_m}\frac{\Gamma}{H}.
\label{eq:fv}
\end{equation}
We compare~$f_v\sigma_8(z)$ with the measurements from 6dFGS~\cite{Beutler:2012px}, SDSS MGS~\cite{Howlett:2014opa}, BOSS DR12 and eBOSS DR16~\cite{eBOSS:2020yzd}, which are not included in the likelihood.
For~$\Gamma < 0$ the second term in eq.~\eqref{eq:fv} is positive and partly compensates the lower~$\sigma_8(z)$ at~$z \lesssim 0.3$.
Using the diagonal errors, the seven measurements give~$\chi^2 = 5.6$--$6.9$ for all models and both data combinations, and the models differ by at most~$1.1$.
Current growth-rate measurements do not discriminate between the models.

\section{Confronting cosmic shear}
\label{sec:shear}

The parameter~$S_8$ combines the present-day values of~$\sigma_8$ and~$\Omega_m$.
Cosmic shear, however, measures the matter distribution integrated along the line of sight, and the values of~$S_8$ quoted by shear surveys are inferred assuming that~$\rho_m \propto (1+z)^3$.
With the exchange this scaling no longer holds at low redshift (section~\ref{sec:model:background}), and the lower~$S_8$ found in section~\ref{sec:results} need not imply a weaker shear signal.
In this section we first compare the models through an effective lensing amplitude as a function of redshift, and then compute the shear correlation functions of each model directly and compare them with the KiDS-Legacy measurements.

\subsection{Effective lensing amplitude}
\label{sec:shear:S8}

The lensing kernel is proportional to the matter density, and the shear power spectrum to the square of the amplitude of the matter fluctuations.
We therefore define the effective lensing amplitude
\begin{equation}
Y(z) \equiv \sigma_8(z)\sqrt{\frac{\Omega_m^{\rm eff}(z)}{0.3}},
\label{eq:Y}
\end{equation}
where~$\sigma_8(z)$ is the linear rms fluctuation in spheres of radius~$8\,h^{-1}$Mpc at redshift~$z$ and~$\Omega_m^{\rm eff}(z)$ is defined in eq.~\eqref{eq:Omeff}.
Today~$Y(0) = S_8$, and in \LCDM{} the ratio~$Y(z)/\sigma_8(z)$ is constant.
We compute~$Y(z)$ for~$60$ samples drawn from each posterior, independently of the samples used for the shear predictions of section~\ref{sec:shear:kids}.
For reference, the lensing kernels of the six KiDS-Legacy source bins peak between~$z \simeq 0.15$ and~$0.6$.

\begin{figure}[t]
\centering
\includegraphics[width=0.9\textwidth]{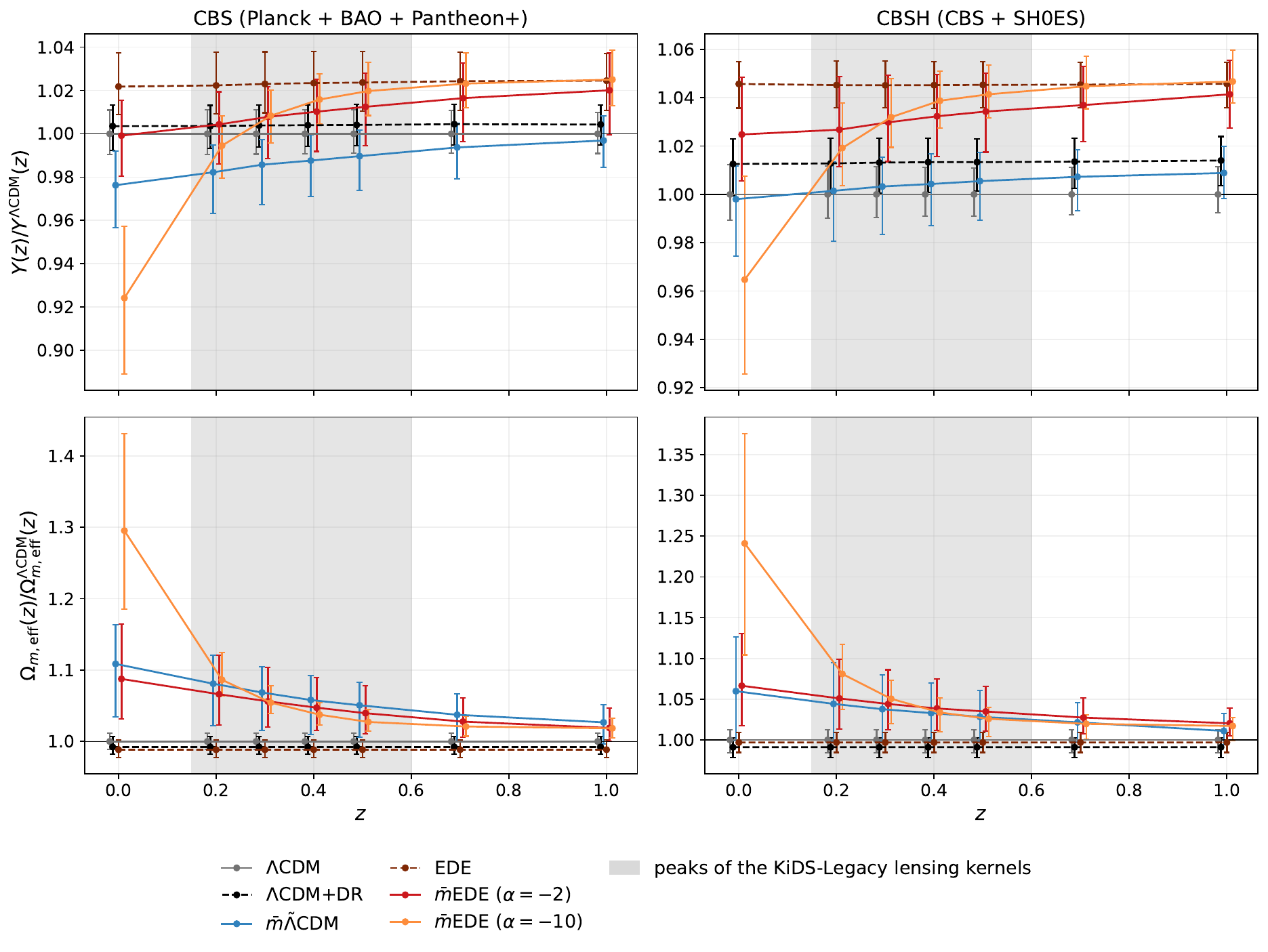}
\caption{Effective lensing amplitude~$Y(z)$, eq.~\eqref{eq:Y} (top), and~$\Omega_m^{\rm eff}(z)$ (bottom), relative to the median of \LCDM{} for the same data, for CBS (left) and CBSH (right).
Points and error bars show the median and the~$68\%$ range of~$60$ posterior samples; the points are slightly offset in redshift for clarity.
The models without exchange are shown with dashed lines.
The grey band marks the range in which the lensing kernels of the KiDS-Legacy source bins peak.}
\label{fig:Y}
\end{figure}

Figure~\ref{fig:Y} shows~$Y(z)$ and~$\Omega_m^{\rm eff}(z)$ relative to \LCDM.
For the models without exchange the ratios are almost independent of redshift: \LCDM+DR is~$0.4\%$ (CBS) and~$1.3\%$ (CBSH) above \LCDM, and EDE~$2.2\%$ and~$4.5\%$ above.
With the exchange the reduction is largest today and decreases with redshift.
For \mbarLCDM{} relative to \LCDM+DR, $Y$ is lower by~$2.7\%$ (CBS) and~$1.4\%$ (CBSH) at~$z = 0$, by~$1.8\%$ and~$1.0\%$ at~$z = 0.3$, and by~$0.7\%$ and~$0.5\%$ at~$z = 1$.
For \mbarEDE{} with~$\alpha = -2$ the reduction relative to EDE is~$2\%$ at~$z = 0$, $1.5\%$ at~$z = 0.3$ and~$0.4\%$ at~$z = 1$ for both data combinations.
For~$\alpha = -10$, $Y(0)$ is lower than in EDE by~$10\%$ (CBS) and~$8\%$ (CBSH), but the difference falls to~$1.2$--$1.5\%$ at~$z = 0.3$, and at~$z \ge 0.5$ the two models agree within~$0.5\%$.
The matter added by the vacuum is present only at late times, so~$\Omega_m^{\rm eff}(z)$ approaches the value without exchange at higher redshift, and the growth of structure is suppressed only after the exchange becomes active.
The reduction of~$S_8$ therefore overstates the change of the lensing amplitude at the redshifts that dominate the shear signal, in particular for~$\alpha = -10$.

\subsection{KiDS-Legacy shear correlation functions}
\label{sec:shear:kids}

\subsubsection{Method}
\label{sec:shear:method}

We use the shear two-point correlation functions~$\xi_\pm(\theta)$ of KiDS-Legacy in six tomographic bins~\cite{Wright:2025xka}, with the covariance matrix of ref.~\cite{Reischke:2024fvk}.
We apply the scale cuts of the KiDS-Legacy analysis, $\theta > 2'$ for~$\xi_+$ and~$\theta > 4'$ for~$\xi_-$, which leave~$357$ of the~$378$ data points.
In the Limber approximation~\cite{LoVerde:2008re} the shear power spectrum of the bins~$i$ and~$j$ is
\begin{equation}
C_\ell^{ij} = \int_0^{r_{\rm H}} \frac{q_i(r)\, q_j(r)}{r^2}\, P_{\rm nl}\!\left(k = \frac{\ell + 1/2}{r}, z(r)\right) \mathrm{d}r ,
\end{equation}
where~$r$ is the comoving distance and~$P_{\rm nl}$ the non-linear matter power spectrum.
We compute the lensing kernel with the matter density of each model,
\begin{equation}
q_i(r) = \frac{3}{2} H_0^2\, \Omega_m^{\rm eff}(z)\,(1+z)\, r \int_r^{r_{\rm H}} n_i(r')\, \frac{r' - r}{r'}\, \mathrm{d}r' ,
\label{eq:qkernel}
\end{equation}
where~$n_i$ is the normalised redshift distribution of bin~$i$; in \LCDM{} $\Omega_m^{\rm eff}(z) = \Omega_m$ and eq.~\eqref{eq:qkernel} reduces to the standard kernel.
We compute~$P_{\rm nl}$ with HMcode-2020~\cite{Mead:2020vgs} without baryonic feedback, applied to the linear power spectrum of each model.
Intrinsic alignments are described by the non-linear alignment model~\cite{Hirata:2004gc,Bridle:2007ft}, with the intrinsic-alignment kernel proportional to~$-A_{\rm IA} C_1\rho_{\rm crit}\,\Omega_m^{\rm eff}(z)/D(z)$, where~$C_1\rho_{\rm crit} = 0.0134$ and~$D$ is the linear growth factor normalised to unity today; we use~$\Omega_m^{\rm eff}(z)$ for consistency with eq.~\eqref{eq:qkernel}, and in \LCDM{} the kernel reduces to the standard form.
The correlation functions follow from~$\xi_\pm(\theta) = \int \ell\, \mathrm{d}\ell/(2\pi)\, C_\ell\, J_{0/4}(\ell\theta)$, evaluated at the effective angular separations of the data.

For each model and data combination we draw~$60$ samples from the posterior and compute the~$\chi^2$ of the KiDS-Legacy data for each of them.
The amplitude~$A_{\rm IA}$ is the only nuisance parameter; we minimise~$\chi^2$ over~$A_{\rm IA} \in [-2, 4]$ for each sample, and keep the redshift distributions and the shear calibration fixed.
We quote the mean~$\chi^2$ over the samples, which measures how well the predictions of a posterior agree with the shear data; since it includes the scatter of the predictions, it penalises the models with exchange, whose posteriors are broader, and compare models through~$\Delta\chi^2$ relative to \LCDM{} and to the corresponding model without exchange.
The shear data are not included in the sampling: the cosmic-shear likelihoods available to us compute the lensing kernel with~$\rho_m \propto (1+z)^3$, which does not hold for the models with exchange.
We check the calculation for \LCDM{} in appendix~\ref{app:shear}.

\subsubsection{Results}
\label{sec:shear:results}

\begin{figure}[t]
\centering
\includegraphics[width=\textwidth]{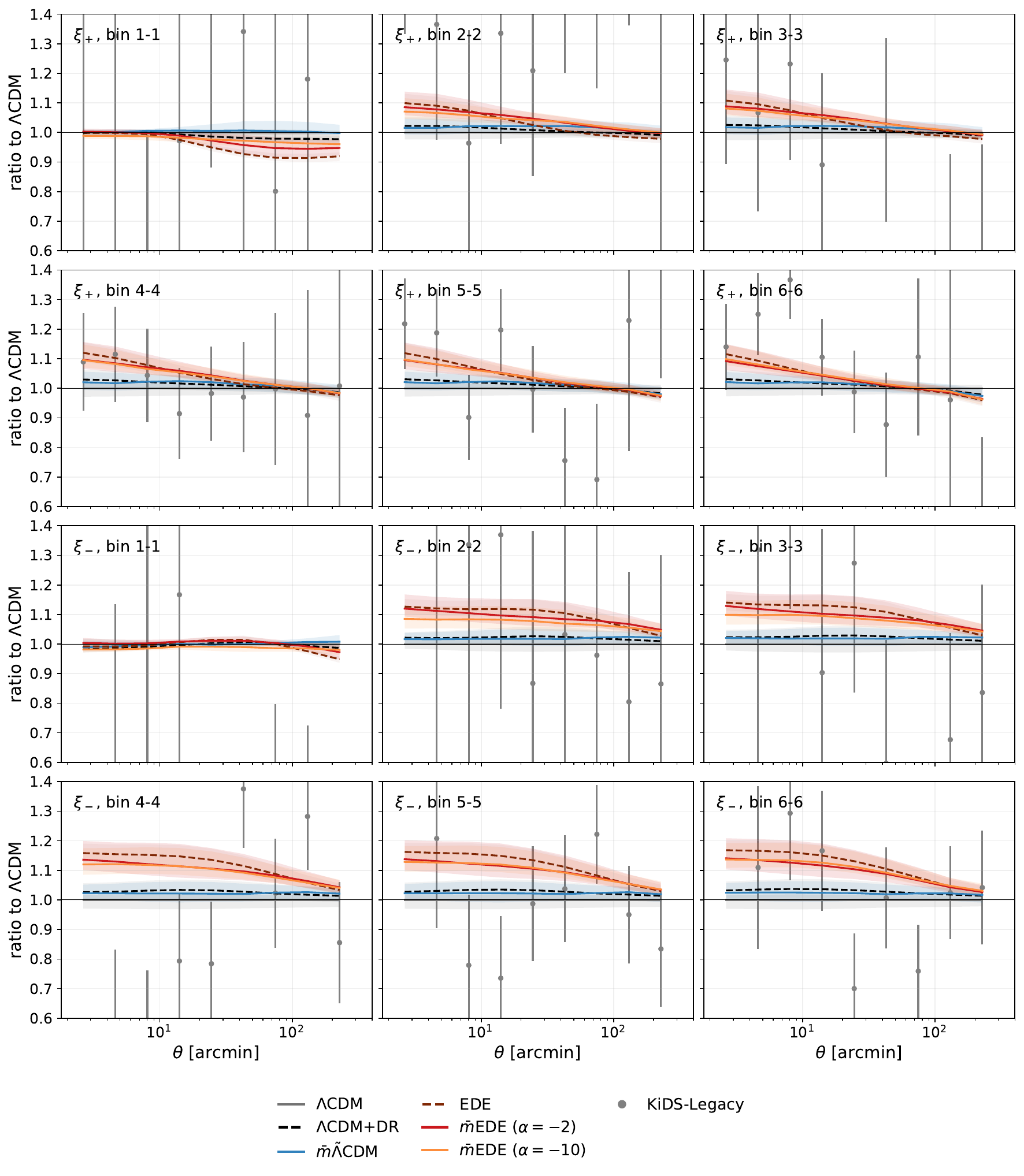}
\caption{KiDS-Legacy correlation functions~$\xi_+$ (upper two rows) and~$\xi_-$ (lower two rows) for the auto-correlations of the six tomographic bins, relative to the median prediction of \LCDM{} for CBSH.
Lines and bands show the median and the~$68\%$ range of the predictions of~$60$ posterior samples for CBSH; the models without exchange are shown with dashed lines.
The points show the data with the square root of the diagonal of the covariance matrix; points outside the scale cuts are not shown.
The corresponding figure for CBS is shown in appendix~\ref{app:shear}.}
\label{fig:shear}
\end{figure}

Table~\ref{tab:shear} lists the results, and figure~\ref{fig:shear} compares the predicted~$\xi_\pm$ with the data for the auto-correlations of the six bins.
In the dark-radiation models the~$\chi^2$ of the shear data changes by less than~$3$ relative to \LCDM, and the exchange lowers it by~$2.0$ (CBS) and~$0.8$ (CBSH) relative to \LCDM+DR.
EDE raises the~$\chi^2$ by~$12.2$ for CBS and~$22.7$ for CBSH.
For CBSH the predicted~$\xi_\pm$ of EDE lie about~$10$--$15\%$ above \LCDM{} on scales~$\theta \lesssim 30'$ in bins~$2$ to~$6$, which reflects the higher lensing amplitude and the larger~$n_s$.
The fit also prefers a larger intrinsic-alignment amplitude, $A_{\rm IA} \simeq 0.45$--$0.5$ compared with~$0.2$--$0.3$ in \LCDM, which partly compensates the higher amplitude.
For CBS and CBSH, respectively, the exchange reduces this excess by~$3.6$ and~$3.0$ for~$\alpha = -2$ and by~$10.8$ and~$5.6$ for~$\alpha = -10$, but does not remove it: relative to \LCDM{} the~$\chi^2$ remains higher by~$8.6$ and~$19.7$ for~$\alpha = -2$, and by~$1.4$ and~$17.1$ for~$\alpha = -10$.
The exchange lowers~$Y(z)$ mainly at~$z \lesssim 0.3$, where the kernels of the lower source bins peak, whereas at~$z \gtrsim 0.5$ the amplitude of \mbarEDE{} with~$\alpha = -10$ equals that of EDE (section~\ref{sec:shear:S8}), so the excess in the higher source bins remains.
For CBS, where EDE has a smaller~$f_{\rm EDE}$ and a smaller excess in~$Y$, the~$\chi^2$ of \mbarEDE{} with~$\alpha = -10$ is only~$1.4$ above \LCDM.

The~$\chi^2$ varies between posterior samples with a standard deviation of~$1$--$3$ for the \LCDM{} and dark-radiation models and~$6$--$12$ for the EDE models, so the statistical uncertainty of the mean~$\chi^2$ from~$60$ samples is between~$0.1$ and~$1.5$.
The differences between EDE and \mbarEDE{} are larger than this uncertainty, while those of order unity among the dark-radiation models are not significant.

\begin{table}[t]
\centering
\renewcommand{\arraystretch}{1.15}
\caption{Comparison with the KiDS-Legacy correlation functions ($357$ data points).
$\Delta\chi^2$ is the mean over~$60$ posterior samples relative to \LCDM, $\Delta\chi^2_{\rm ex}$ is relative to the corresponding model without exchange, and~$A_{\rm IA}$ is the mean of the best-fitting intrinsic-alignment amplitude.}
\label{tab:shear}
\begin{tabular}{lcccccc}
\hline
 & \multicolumn{3}{c}{CBS} & \multicolumn{3}{c}{CBSH} \\
Model & $\Delta\chi^2$ & $\Delta\chi^2_{\rm ex}$ & $A_{\rm IA}$ & $\Delta\chi^2$ & $\Delta\chi^2_{\rm ex}$ & $A_{\rm IA}$ \\
\hline
\LCDM & $0$ & -- & $0.30$ & $0$ & -- & $0.23$ \\
\LCDM+DR & $+1.2$ & -- & $0.31$ & $+2.7$ & -- & $0.31$ \\
\mbarLCDM & $-0.8$ & $-2.0$ & $0.26$ & $+1.9$ & $-0.8$ & $0.29$ \\
EDE & $+12.2$ & -- & $0.44$ & $+22.7$ & -- & $0.50$ \\
\mbarEDE{} ($\alpha = -2$) & $+8.6$ & $-3.6$ & $0.41$ & $+19.7$ & $-3.0$ & $0.48$ \\
\mbarEDE{} ($\alpha = -10$) & $+1.4$ & $-10.8$ & $0.32$ & $+17.1$ & $-5.6$ & $0.48$ \\
\hline
\end{tabular}
\end{table}

\subsubsection{Robustness and limitations}
\label{sec:shear:robust}

Our comparison is simpler than the KiDS-Legacy analysis in several respects.
First, we keep the redshift distributions, the shear calibration and the baryonic feedback fixed, and the only nuisance parameter is~$A_{\rm IA}$.
Marginalising over these parameters would reduce the absolute differences in~$\chi^2$, in particular for the EDE models, whose excess is concentrated on small scales where baryonic feedback suppresses the power.
Since the dark-radiation models and the models with and without exchange differ mainly in the overall amplitude, we expect the ordering of the models to be less sensitive to these choices than the size of the differences.
Second, the fiducial KiDS-Legacy constraints are based on COSEBIs and band powers~\cite{Wright:2025xka}; the correlation functions are provided for comparison, and we evaluate the predictions at the effective angular separations without averaging over the angular bins.
Third, HMcode is calibrated on $N$-body simulations without energy exchange~\cite{Mead:2020vgs}.
We apply it to the linear power spectrum of each model, which captures the change in the amplitude of the linear fluctuations, but not the changes in the abundance and internal structure of haloes caused by the inflow of homogeneous dark matter, which modifies spherical collapse and the virial overdensity~\cite{Wang:2016lxa}.
The size of this effect depends on how much dark matter is transferred while haloes form: about~$5$--$7\%$ of the present CDM density at~$z < 0.7$ for~$\alpha = -2$, and~$20$--$25\%$ for~$\alpha = -10$.
For~$\alpha = -2$ we expect the error to be small compared with the differences between the models, whereas for~$\alpha = -10$ the predictions on small scales, which dominate~$\xi_-$, should be regarded as indicative; simulations of an interacting vacuum with geodesic CDM would be needed to model these scales reliably.
Finally, the comparison is a test of the posterior predictions of each model rather than a joint fit, and we do not include other shear surveys such as DES~Y3.
A joint analysis would require a shear likelihood with the lensing kernel of eq.~\eqref{eq:qkernel} and the full set of nuisance parameters, which we leave to future work.

\section{Discussion}
\label{sec:discussion}

\subsection{The exchange with different early-time solutions}
\label{sec:discussion:EDE}

The exchange acts in the same way on both early-time solutions.
For CBSH and~$\alpha = -2$ it lowers~$S_8$ by~$0.014$ in the dark-radiation model and by~$0.013$ in the EDE model, and changes~$H_0$ by at most~$0.3~\mathrm{km\,s^{-1}\,Mpc^{-1}}$ in both.
The early-time solutions, however, raise~$S_8$ by different amounts.
Dark radiation raises~$S_8$ by~$0.011$ relative to \LCDM, and the exchange removes this increase.
EDE raises~$S_8$ by~$0.036$, through a larger~$\omega_{\rm cdm}$ and a larger~$n_s$, and the exchange with~$\alpha = -2$ removes only about a third of it.
For~$\alpha = -10$ the decrease of~$S_8$ exceeds the increase due to EDE, but the decrease of the lensing amplitude is concentrated at~$z \lesssim 0.3$, and at the redshifts probed by the higher source bins of KiDS-Legacy the lensing amplitude remains that of EDE (section~\ref{sec:shear}).
The exchange changes the amplitude of the matter fluctuations at late times, whereas the larger~$n_s$ of EDE also changes the shape of the power spectrum, which a late-time exchange cannot compensate.
The exchange is therefore most effective when the early-time solution raises~$S_8$ mainly through~$\omega_{\rm cdm}$, as dark radiation does.

The idea of combining a component that raises~$H_0$ with a component that suppresses the growth of structure was explored in ref.~\cite{Allali:2021azp}, where an ultralight axion suppresses the matter power spectrum on scales below its Jeans length.
In our case the suppression comes from the late-time inflow of dark matter without density perturbations, eq.~\eqref{eq:deltac}.
Since the extra term in eq.~\eqref{eq:deltac} does not depend on~$k$, the suppression is nearly scale-independent on sub-horizon scales and is confined to low redshift, and it is accompanied by an increase of the matter density, $\Omega_m^{\rm eff}(z)$, that is absent in the axion case.
The two mechanisms leave different imprints on the scale and redshift dependence of the clustering amplitude, which tomographic shear and galaxy clustering data could distinguish.

\subsection{Physical interpretation of the preferred coupling}
\label{sec:discussion:coupling}

In all cases the posteriors favour~$\bar m < 0$, i.e.\ an energy flow from the vacuum into the dark sector.
This is the direction expected from particle production by the vacuum in an expanding universe~\cite{Parker:1973qd}, which motivates the $\tilde\Lambda$CDM model~\cite{Xue:2014kna,Xue:2023qft}.
The flow in the opposite direction is not favoured: with~$\bar m \ge 0$ the posterior peaks at zero and the fit is slightly worse than without exchange (section~\ref{sec:results:alpha}).
The preference is stronger for CBS than for CBSH, and the coupling is correlated mainly with~$\sigma_8$ and~$\Omega_m$ and only weakly with~$H_0$ (section~\ref{sec:results:coupling}), so it is not driven by the SH0ES calibration.
The same direction of the energy flow is found in late-time analyses of interacting dark energy with~$w_{\rm de} = -1$ and~$w_{\rm dm} = 0$, where the preference weakens when~$w_{\rm de}$ is left free and moves to~$w_{\rm de} > -1$~\cite{Wu:2025vrl}; since the exchange itself gives the vacuum an effective equation of state above~$-1$ (see below), the two descriptions are partly degenerate.

The preferred coupling corresponds to a present rate~$\Gamma/H \simeq -0.1$ to~$-0.15$ for~$\alpha = -2$, so the vacuum transfers about~$10$--$15\%$ of the CDM energy per $e$-fold today.
Since the vacuum loses this energy, its density decreases slowly, and its effective equation of state obtained from eqs.~\eqref{eq:rhoL}--\eqref{eq:rhodr} is
\begin{equation}
w_{\rm eff} \equiv -1 - \frac{\dot\rho_\Lambda}{3H\rho_\Lambda} = -1 - \frac{\Gamma}{3H}\,\frac{\rho_{\rm c} + \rho_{\rm dr}}{\rho_\Lambda},
\end{equation}
which for~$\Gamma < 0$ lies above~$-1$.
Since~$w_{\rm eff}$ stays above~$-1$ at all times, the exchange does not produce the crossing of~$w = -1$ preferred by the DESI DR2 analysis of the~$w_0w_a$ parametrisation~\cite{DESI:2025zgx}.
With~$\rho_{\rm c}/\rho_\Lambda \simeq 0.4$ today, $w_{\rm eff} + 1 \simeq 0.015$--$0.02$ for~$\alpha = -2$.
For~$\alpha = -10$ the present rate is~$\Gamma/H \simeq -1.2$ to~$-1.9$, and~$w_{\rm eff} + 1 \simeq 0.2$--$0.35$ today, but the exchange switches on only at~$z \lesssim 0.7$.
In this case~$\Omega_m^{\rm eff}$ relative to \LCDM{} increases by about~$20$--$30\%$ between~$z \simeq 0.7$ and today (figure~\ref{fig:Y}), most of it at~$z \lesssim 0.2$.
For~$\alpha \ge -1$ the exchange is active over a wider range of redshift, and the data restrict the coupling to values for which the effect on~$S_8$ is small (section~\ref{sec:results:alpha}).
An exchange that switches on late, with~$\alpha \lesssim -2$, therefore allows a larger reduction of~$S_8$, although the differences in~$\chi^2$ between the values of~$\alpha$ are small.

\subsection{Caveats}
\label{sec:discussion:caveats}

Our analysis has several limitations.
We use the one-parameter EDE model with~$\log_{10} a_c$ and~$\theta_i$ fixed~\cite{Murgia:2020ryi,Smith:2020rxx}.
Varying these parameters broadens the EDE posterior, but the late-time exchange acts in the same way for any early-time solution, and we expect the shift in~$S_8$ caused by the exchange to be similar.
We treat~$\alpha = -10$ as a limiting case that confines the exchange to~$z \lesssim 0.7$, in order to show how the effect on~$S_8$ depends on when the exchange is active, rather than as a model preferred on physical grounds.
In this case the present rate is large, $|\Gamma/H| \simeq 1$--$2$, and about~$20$--$25\%$ of the present CDM density is transferred from the vacuum at~$z \lesssim 0.7$, compared with~$7$--$10\%$ over the whole history for~$\alpha = -2$.
The linear equations remain valid for any value of~$\Gamma$, since the energy is transferred along the CDM four-velocity and the vacuum, with~$w = -1$, does not cluster, so that the early-time instability of interacting dark-energy fluids does not arise~\cite{Wands:2012vg,Wang:2014xca}; the large change of~$\Omega_m$ at low redshift, however, makes the non-linear modelling discussed below less reliable for this case.
The significance of the coupling depends on the prior on~$\bar m$ in the usual way for a parameter that is consistent with zero at the~$1$--$3\sigma$ level, and the improvement in~$\chi^2$ is estimated from the chains rather than with a minimiser, so differences of order unity in~$\Delta\chi^2$ and~$\Delta{\rm AIC}$ should not be over-interpreted.
We use the Planck 2018 CMB data; ground-based CMB measurements constrain EDE more tightly~\cite{AtacamaCosmologyTelescope:2025nti}, which would likely reduce the increase of~$H_0$ and of~$S_8$ in the EDE models.
The non-linear power spectrum is computed with HMcode, which is calibrated on simulations without energy exchange; this affects the Planck lensing likelihood only weakly, but limits the accuracy of the cosmic-shear predictions on small scales, in particular for~$\alpha = -10$ (section~\ref{sec:shear:robust}).
Finally, our comparison with cosmic shear fixes the redshift distributions, the shear calibration and the baryonic feedback (section~\ref{sec:shear:robust}); a joint analysis of CMB, BAO, SNe~Ia and shear data with a consistent treatment of the lensing kernel would test whether the preference for~$\bar m < 0$ survives when the shear data enter the likelihood.

\section{Conclusions}
\label{sec:conclusions}

Extensions of \LCDM{} that raise~$H_0$ through additional energy before recombination also raise~$S_8$, because the CMB then requires a larger dark matter density.
We have studied whether an energy exchange between the vacuum and the dark sector, active only at late times, can offset this increase.
The coupling rate scales as~$\bar m H^{\alpha}$, so that for~$\alpha \le -2$ the exchange is negligible before recombination and leaves the sound horizon unchanged.
The energy is transferred along the CDM four-velocity and is unperturbed in the CDM rest frame, so that CDM follows geodesics and the vacuum does not cluster.
When the vacuum decays into CDM, the new dark matter carries no density perturbation: $\Omega_m$ increases, $\sigma_8$ decreases, and~$S_8$ decreases.
We added this exchange to two early-time solutions, dark radiation (\mbarLCDM) and a one-parameter EDE model (\mbarEDE), and constrained the models with Planck 2018, DESI DR2 BAO and Pantheon+ data (CBS), and with the SH0ES calibration added (CBSH).
Our main results are the following.

\begin{itemize}
\item Without the SH0ES calibration the posteriors favour an energy flow from the vacuum into the dark sector, $\bar m < 0$, at~$1.8\sigma$ for \mbarLCDM{} and at~$1.8\sigma$ and~$2.7\sigma$ for \mbarEDE{} with~$\alpha = -2$ and~$-10$, respectively, with~$\Delta\chi^2 = -3.7$, $-1.9$ and~$-7.2$ relative to the models without exchange.
The exchange lowers~$S_8$ from~$0.815$ to~$0.792$ in the dark-radiation model and from~$0.830$ to~$0.812$ and~$0.750$ in the EDE model.
It does not raise~$H_0$, which decreases by~$0.4$--$1.0~\mathrm{km\,s^{-1}\,Mpc^{-1}}$; \mbarLCDM{} gives~$H_0 = 68.83 \pm 0.91~\mathrm{km\,s^{-1}\,Mpc^{-1}}$, $3.1\sigma$ below SH0ES, and \mbarEDE{} lies~$1.7$--$2.0\sigma$ below.

\item With the SH0ES calibration, which holds~$H_0$ at a high value, the exchange lowers~$S_8$ at almost fixed~$H_0$.
\mbarLCDM{} gives~$H_0 = 71.34 \pm 0.70~\mathrm{km\,s^{-1}\,Mpc^{-1}}$ and~$S_8 = 0.800 \pm 0.016$; the corresponding values are~$71.59 \pm 0.68~\mathrm{km\,s^{-1}\,Mpc^{-1}}$ and~$0.815 \pm 0.008$ for \LCDM+DR, and~$68.77 \pm 0.29~\mathrm{km\,s^{-1}\,Mpc^{-1}}$ and~$0.804 \pm 0.008$ for \LCDM.
In the EDE model the exchange lowers~$S_8$ from~$0.840$ to~$0.827$ and~$0.780$ for~$\alpha = -2$ and~$-10$, respectively, with~$H_0$ decreasing by at most~$0.4~\mathrm{km\,s^{-1}\,Mpc^{-1}}$.
The coupling differs from zero by~$1.1$--$2.0\sigma$.
Since these posteriors include SH0ES, the high values of~$H_0$ reflect the calibration and test the exchange conditionally on it, rather than showing that the Hubble tension is eased.

\item The direction and the timing of the exchange matter.
Restricting the flow to the opposite direction raises~$S_8$ and does not improve the fit.
For~$\alpha = 0$ and~$-1$ the exchange is active over a wider range of redshift, the coupling is more tightly constrained, and the reduction of~$S_8$ is smaller; $\alpha = -10$ is a limiting case in which the exchange is confined to~$z \lesssim 0.7$.

\item The reduction of~$S_8$ overstates the effect on cosmic shear.
Since the exchange changes the matter density at low redshift, the reduction of the lensing amplitude is largest today and amounts to only~$1$--$2\%$ at~$z \simeq 0.15$--$0.6$, where the KiDS-Legacy lensing kernels peak.
In the dark-radiation models the~$\chi^2$ of the KiDS-Legacy correlation functions changes by less than~$3$.
EDE raises it by~$12$ (CBS) and~$23$ (CBSH) relative to \LCDM, and the exchange reduces this excess by~$3$--$4$ for~$\alpha = -2$ and by~$6$--$11$ for~$\alpha = -10$, without removing it.
\end{itemize}

A late-time exchange between the vacuum and dark matter can thus offset the increase of~$S_8$ caused by dark radiation and part of that caused by EDE, without undoing the increase of~$H_0$ provided by the early-time solution.
It does not ease the Hubble tension itself, its effect is concentrated at the lowest redshifts, where current shear data have little weight, and it leaves the larger~$n_s$ of EDE unchanged.
The coupling differs from zero by at most~$2.7\sigma$, so current data do not require the exchange.
A joint analysis with a cosmic-shear likelihood that uses the matter density of each model in the lensing kernel would test the preference for~$\bar m < 0$ directly.
Tomographic shear and clustering measurements from Euclid~\cite{Euclid:2024yrr} and LSST~\cite{LSST:2008ijt} will measure the lensing amplitude as a function of redshift, and can distinguish a late-time exchange, whose effect grows towards~$z = 0$, from mechanisms that suppress the growth of structure at all redshifts.
Observations of the redshifted 21-cm line provide a complementary probe.
Intensity mapping of neutral hydrogen with the SKA~\cite{SKA:2018ckk} and other radio telescopes will measure the expansion rate and the growth of structure over~$0 \lesssim z \lesssim 3$~\cite{Bull:2014rha,Wu:2021vfz} and can constrain energy exchange in the dark sector~\cite{Zhang:2021yof}; since the exchange changes both~$\rho_m(z)$ and the growth rate at~$z \lesssim 1$, such data combined with cosmic shear would separate its effects on the background and on the clustering amplitude.

\appendix

\section{Implementation in CLASS and numerical validation}
\label{app:class}

We implement the exchange in version~3.3 of \texttt{CLASS}.
The background equations~\eqref{eq:rhoL}--\eqref{eq:rhodr} are integrated together with the other background quantities.
The input parameters are the present-day densities~$\omega_{\rm cdm}$ and~$N_{\rm dr}$, and the present vacuum density follows from the flatness condition.
Since the exchange changes the CDM and dark-radiation densities between the initial time and today, the initial values are not known in advance.
We obtain them by integrating eqs.~\eqref{eq:rhoL}--\eqref{eq:rhodr} backwards from today to the initial scale factor of \texttt{CLASS} with a fourth-order Runge--Kutta scheme in~$\ln a$, and then integrate forwards with the standard background integrator.
At the best-fitting point of \mbarLCDM{} for CBSH, $\bar m/H_0 = -1845$, the forward integration recovers the input values of~$\omega_{\rm cdm}$ and~$N_{\rm dr}$ to relative precision~$3 \times 10^{-8}$ and~$2 \times 10^{-7}$, and~$H(z = 0)$ agrees with~$H_0$ to~$5 \times 10^{-9}$.
For~$\alpha = 0$ the exchange is switched off at~$z > 10$ (section~\ref{sec:model:coupling}).
The dark radiation is implemented as the interacting dark radiation species of \texttt{CLASS} with fluid perturbations, so that its background density can exchange energy with the vacuum.

In the perturbation module we add the term~$a\Gamma\delta_{\rm c}$ to the CDM continuity equation, eq.~\eqref{eq:deltac}, and keep~$\theta_{\rm c} = 0$ in the synchronous gauge.
The perturbations of the dark-radiation fluid follow the standard fluid equations.
Adding the corresponding exchange terms, $a\Gamma\delta_{\rm dr}$ and~$a\Gamma\theta_{\rm dr}$, to the dark-radiation equations changes the CMB spectra at the best-fitting points for~$\alpha = -2$ and~$-1$ by less than~$4 \times 10^{-6}$ in TT, $2 \times 10^{-6}$ in EE and~$2 \times 10^{-7}$ in the lensing potential, and~$\sigma_8$ by~$10^{-9}$, so these terms have no effect on our results.

We have checked the implementation in several ways.
For~$\bar m = 0$ the modified code reproduces the unmodified \texttt{CLASS}: the~$\chi^2$ of the \LCDM{} best-fitting point is recovered exactly, and \LCDM+DR agrees with the interacting dark radiation fluid of \texttt{CLASS} to~$3 \times 10^{-5}$.
For the EDE models we add the same equations to \texttt{AxiCLASS}.
With~$\bar m = 0$ the modified \texttt{AxiCLASS} reproduces the~$\chi^2$ of the original code at the EDE best-fitting point to all printed digits.
Without EDE and with~$\bar m/H_0 = -1500$, the two implementations give the same change in~$\sigma_8$, $-9.8136\%$, the same factor~$F(a_*) = 0.92903$ at~$z = 1100$, and the same ratio~$\rho_\Lambda(z = 1100)/\rho_\Lambda(0) = 1.19454$, to five significant digits.
Finally, the values of~$\sigma_8$ and~$H_0$ recomputed for the posterior samples used in sections~\ref{sec:results:fs8} and~\ref{sec:shear} agree with those stored in the chains to~$10^{-5}$.

\section{Cosmic shear pipeline validation}
\label{app:shear}

The predictions of section~\ref{sec:shear:kids} use a simplified pipeline with~$A_{\rm IA}$ as the only nuisance parameter.
To assess it, we have also implemented the full KiDS-Legacy correlation-function likelihood following the official pipeline~\cite{Wright:2025xka}: the theory is averaged over the angular bins with the measured pair counts, the redshift distributions are shifted with the correlated priors of the redshift calibration, the intrinsic alignment follows the mass-dependent model of the KiDS-Legacy analysis, and the non-linear power spectrum includes baryonic feedback through HMcode-2020.
For~$30$ samples drawn at random from the official KiDS-Legacy chain, this likelihood reproduces~$\sigma_8$ to within~$0.3\%$ and the~$\chi^2$ of the official analysis with a mean difference of~$-1.3$ and a scatter of~$1.9$, for~$\chi^2 \simeq 410$.
In this comparison the AGN heating temperature of HMcode was fixed to $\log_{10}(T_{\rm AGN}/{\rm K})=8.0$ rather than taken from the official chain, which accounts for most of the remaining scatter.

\begin{figure}[t]
\centering
\includegraphics[width=\textwidth]{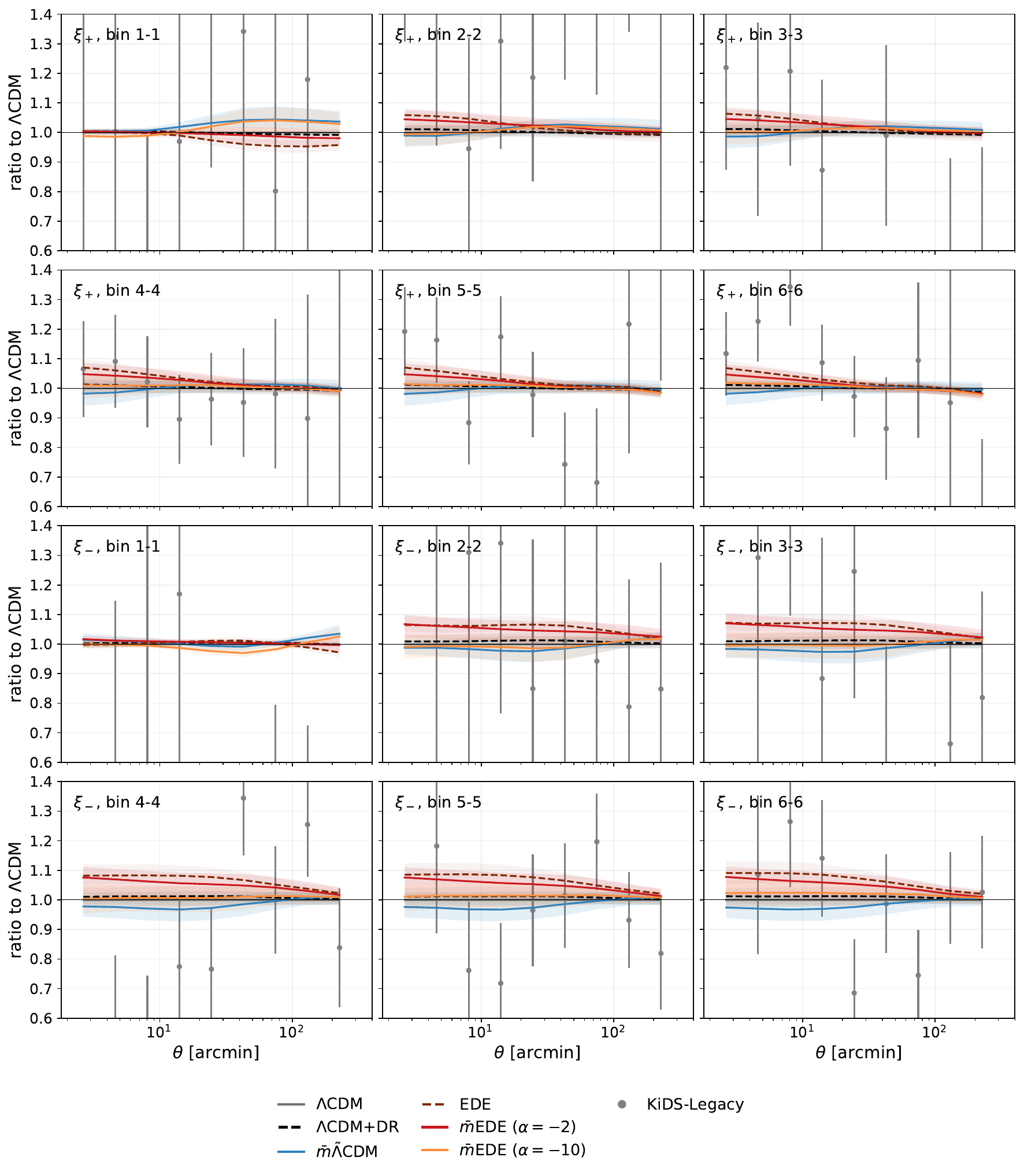}
\caption{As figure~\ref{fig:shear}, for CBS.}
\label{fig:shear_CBS}
\end{figure}

With the nuisance parameters fixed, the simplified pipeline gives larger values of~$\chi^2$: for the \LCDM{} posteriors the mean is~$430$ (CBS) and~$428$ (CBSH) for~$357$ data points.
We therefore interpret only the differences in~$\chi^2$ between models, which are computed with the same fixed nuisance parameters.
Rescaling the predicted~$\xi_\pm$ of \LCDM{} by a free amplitude gives a best-fitting factor of~$0.95 \pm 0.02$ for CBS and~$0.98 \pm 0.02$ for CBSH, consistent with the agreement between the value of~$S_8$ found by KiDS-Legacy~\cite{Wright:2025xka} and that inferred from the CMB in \LCDM{} at the level of a few per cent.
In \LCDM{} the lensing kernel of eq.~\eqref{eq:qkernel} reduces to the standard one, so the differences between models come only from the background and the power spectrum of each model.
Figure~\ref{fig:shear_CBS} shows the predictions for CBS.

\section{Convergence diagnostics}
\label{app:convergence}

During sampling we monitored the Gelman--Rubin statistic~\cite{Gelman:1992zz} of the cosmological parameters, computed after removing the first~$30\%$ of each chain, and stopped the chains when~$R - 1 < 0.02$ in two consecutive checks.
Table~\ref{tab:convergence} lists the final values computed with \texttt{MontePython}, which removes the initial part of each chain before the sampler reaches the region of high likelihood.
Twelve of the fifteen runs have~$R - 1 < 0.02$, and all have~$R - 1 < 0.025$.
These values are below the thresholds commonly used in analyses of EDE models~\cite{Murgia:2020ryi,Smith:2020rxx}.

\begin{table}[t]
\centering
\renewcommand{\arraystretch}{1.15}
\caption{Convergence of the chains: the largest Gelman--Rubin~$R - 1$ among the cosmological parameters, computed by \texttt{MontePython}, the parameter for which it is reached, and the number of steps after burn-in (including repeated points), summed over the four chains.}
\label{tab:convergence}
\begin{tabular}{llccc}
\hline
Data & Model & $R - 1$ & Parameter & Steps \\
\hline
CBS & \LCDM & $0.0110$ & $\omega_b$ & $343{,}454$ \\
CBS & \LCDM+DR & $0.0087$ & $\omega_b$ & $224{,}474$ \\
CBS & \mbarLCDM{} ($\alpha = -2$) & $0.0190$ & $\omega_b$ & $251{,}385$ \\
CBS & EDE & $0.0184$ & $f_{\rm EDE}$ & $364{,}136$ \\
CBS & \mbarEDE{} ($\alpha = -2$) & $0.0131$ & $\tau_{\rm reio}$ & $215{,}223$ \\
CBS & \mbarEDE{} ($\alpha = -10$) & $0.0190$ & $\omega_{\rm cdm}$ & $1{,}211{,}029$ \\
CBSH & \LCDM & $0.0060$ & $\tau_{\rm reio}$ & $348{,}754$ \\
CBSH & \LCDM+DR & $0.0150$ & $\omega_{\rm cdm}$ & $347{,}455$ \\
CBSH & \mbarLCDM{} ($\alpha = -2$) & $0.0131$ & $100\theta_s$ & $241{,}291$ \\
CBSH & EDE & $0.0074$ & $\tau_{\rm reio}$ & $317{,}321$ \\
CBSH & \mbarEDE{} ($\alpha = -2$) & $0.0180$ & $\bar m/H_0$ & $1{,}130{,}608$ \\
CBSH & \mbarEDE{} ($\alpha = -10$) & $0.0237$ & $\omega_b$ & $445{,}857$ \\
CBSH & \mbarLCDM{} ($\alpha = 0$) & $0.0222$ & $n_s$ & $224{,}905$ \\
CBSH & \mbarLCDM{} ($\alpha = -1$) & $0.0200$ & $\omega_{\rm cdm}$ & $292{,}759$ \\
CBSH & \mbarLCDM{} ($\alpha = -2$, $\bar m \ge 0$) & $0.0091$ & $\omega_{\rm cdm}$ & $305{,}780$ \\
\hline
\end{tabular}
\end{table}

\paragraph{Data availability.} The modified versions of \texttt{CLASS} and \texttt{AxiCLASS} and the chains are available from the first author upon reasonable request.

\acknowledgments

This work was supported by the Early-Career Young Scientists and Technologists Project of Jiangxi Province under Grant No.~20262BEJ730225, the National Natural Science Foundation of China under Grant Nos.~12533001, 12473001 and 12575049, the National SKA Program of China under Grant Nos.~2022SKA0110200 and 2022SKA0110203, the China Manned Space Program under Grant No.~CMSCSST2025A02, and the 111 Project under Grant No.~B16009.
This work is based on observations obtained with Planck (\url{http://www.esa.int/Planck}), an ESA science mission with instruments and contributions directly funded by ESA Member States, NASA, and Canada.
We thank the DESI collaboration for making the DR2 BAO measurements publicly available, and the Pantheon+ and SH0ES teams for releasing their data.
The cosmic-shear data are based on observations made with ESO Telescopes at the La Silla Paranal Observatory under programme IDs 179.A-2004, 177.A-3016, 177.A-3017, 177.A-3018 and 298.A-5015.
We used Claude (Anthropic) for language editing of the manuscript; the authors take full responsibility for the content.

\bibliographystyle{JHEP}
\bibliography{biblio.bib}

\providecommand{\href}[2]{#2}\begingroup\raggedright\begin{thebibliography}{10}

\bibitem{Planck:2018vyg}
{\scshape Planck} collaboration, \emph{{Planck 2018 results. VI. Cosmological
  parameters}},
  \href{https://doi.org/10.1051/0004-6361/201833910}{\emph{Astron. Astrophys.}
  {\bfseries 641} (2020) A6}
  [\href{https://arxiv.org/abs/1807.06209}{{\ttfamily 1807.06209}}].

\bibitem{DESI:2025zgx}
{\scshape DESI} collaboration, \emph{{DESI DR2 results. II. Measurements of
  baryon acoustic oscillations and cosmological constraints}},
  \href{https://doi.org/10.1103/tr6y-kpc6}{\emph{Phys. Rev. D} {\bfseries 112}
  (2025) 083515} [\href{https://arxiv.org/abs/2503.14738}{{\ttfamily
  2503.14738}}].

\bibitem{Scolnic:2021amr}
D.~Scolnic et~al., \emph{{The Pantheon+ Analysis: The Full Data Set and
  Light-curve Release}},
  \href{https://doi.org/10.3847/1538-4357/ac8b7a}{\emph{Astrophys. J.}
  {\bfseries 938} (2022) 113}
  [\href{https://arxiv.org/abs/2112.03863}{{\ttfamily 2112.03863}}].

\bibitem{Brout:2022vxf}
D.~Brout et~al., \emph{{The Pantheon+ Analysis: Cosmological Constraints}},
  \href{https://doi.org/10.3847/1538-4357/ac8e04}{\emph{Astrophys. J.}
  {\bfseries 938} (2022) 110}
  [\href{https://arxiv.org/abs/2202.04077}{{\ttfamily 2202.04077}}].

\bibitem{Verde:2019ivm}
L.~Verde, T.~Treu and A.G.~Riess, \emph{{Tensions between the Early and the
  Late Universe}},
  \href{https://doi.org/10.1038/s41550-019-0902-0}{\emph{Nature Astron.}
  {\bfseries 3} (2019) 891} [\href{https://arxiv.org/abs/1907.10625}{{\ttfamily
  1907.10625}}].

\bibitem{DiValentino:2021izs}
E.~Di~Valentino, O.~Mena, S.~Pan, L.~Visinelli, W.~Yang, A.~Melchiorri et~al.,
  \emph{{In the realm of the Hubble tension{\textemdash}a review of
  solutions}}, \href{https://doi.org/10.1088/1361-6382/ac086d}{\emph{Class.
  Quant. Grav.} {\bfseries 38} (2021) 153001}
  [\href{https://arxiv.org/abs/2103.01183}{{\ttfamily 2103.01183}}].

\bibitem{Abdalla:2022yfr}
E.~Abdalla et~al., \emph{{Cosmology intertwined: A review of the particle
  physics, astrophysics, and cosmology associated with the cosmological
  tensions and anomalies}},
  \href{https://doi.org/10.1016/j.jheap.2022.04.002}{\emph{JHEAp} {\bfseries
  34} (2022) 49} [\href{https://arxiv.org/abs/2203.06142}{{\ttfamily
  2203.06142}}].

\bibitem{Li:2026asg}
T.-N.~Li, G.-H.~Du, H.~Wang, Y.-H.~Li, J.-F.~Zhang and X.~Zhang, \emph{{Dark
  Energy in the DESI Era: A Brief Review of Evidence,
  Beyond-{\ensuremath{\Lambda}}CDM Interpretations, and Tensions}},
  \href{https://doi.org/10.1088/1674-4527/ae8429}{\emph{Res. Astron.
  Astrophys.} {\bfseries 26} (2026) 084002}
  [\href{https://arxiv.org/abs/2606.21826}{{\ttfamily 2606.21826}}].

\bibitem{Riess:2021jrx}
A.G.~Riess et~al., \emph{{A Comprehensive Measurement of the Local Value of the
  Hubble Constant with 1 km s$^{-1}$ Mpc$^{-1}$ Uncertainty from the Hubble
  Space Telescope and the SH0ES Team}},
  \href{https://doi.org/10.3847/2041-8213/ac5c5b}{\emph{Astrophys. J. Lett.}
  {\bfseries 934} (2022) L7}
  [\href{https://arxiv.org/abs/2112.04510}{{\ttfamily 2112.04510}}].

\bibitem{Riess:2024ohe}
A.G.~Riess, G.S.~Anand, W.~Yuan, S.~Casertano, A.~Dolphin, L.M.~Macri et~al.,
  \emph{{JWST Observations Reject Unrecognized Crowding of Cepheid Photometry
  as an Explanation for the Hubble Tension at 8{\ensuremath{\sigma}}
  Confidence}},
  \href{https://doi.org/10.3847/2041-8213/ad1ddd}{\emph{Astrophys. J. Lett.}
  {\bfseries 962} (2024) L17}
  [\href{https://arxiv.org/abs/2401.04773}{{\ttfamily 2401.04773}}].

\bibitem{Riess:2024vfa}
A.G.~Riess et~al., \emph{{JWST Validates HST Distance Measurements: Selection
  of Supernova Subsample Explains Differences in JWST Estimates of Local H
  $_{0}$}}, \href{https://doi.org/10.3847/1538-4357/ad8c21}{\emph{Astrophys.
  J.} {\bfseries 977} (2024) 120}
  [\href{https://arxiv.org/abs/2408.11770}{{\ttfamily 2408.11770}}].

\bibitem{Freedman:2024eph}
W.L.~Freedman, B.F.~Madore, T.J.~Hoyt, I.S.~Jang, A.J.~Lee and K.A.~Owens,
  \emph{{Status Report on the Chicago-Carnegie Hubble Program (CCHP):
  Measurement of the Hubble Constant Using the Hubble and James Webb Space
  Telescopes}},
  \href{https://doi.org/10.3847/1538-4357/adce78}{\emph{Astrophys. J.}
  {\bfseries 985} (2025) 203}
  [\href{https://arxiv.org/abs/2408.06153}{{\ttfamily 2408.06153}}].

\bibitem{KiDS:2020suj}
{\scshape KiDS} collaboration, \emph{{KiDS-1000 Cosmology: Cosmic shear
  constraints and comparison between two point statistics}},
  \href{https://doi.org/10.1051/0004-6361/202039070}{\emph{Astron. Astrophys.}
  {\bfseries 645} (2021) A104}
  [\href{https://arxiv.org/abs/2007.15633}{{\ttfamily 2007.15633}}].

\bibitem{DES:2021bvc}
{\scshape DES} collaboration, \emph{{Dark Energy Survey Year 3 results:
  Cosmology from cosmic shear and robustness to data calibration}},
  \href{https://doi.org/10.1103/PhysRevD.105.023514}{\emph{Phys. Rev. D}
  {\bfseries 105} (2022) 023514}
  [\href{https://arxiv.org/abs/2105.13543}{{\ttfamily 2105.13543}}].

\bibitem{DES:2021vln}
{\scshape DES} collaboration, \emph{{Dark Energy Survey Year 3 results:
  Cosmology from cosmic shear and robustness to modeling uncertainty}},
  \href{https://doi.org/10.1103/PhysRevD.105.023515}{\emph{Phys. Rev. D}
  {\bfseries 105} (2022) 023515}
  [\href{https://arxiv.org/abs/2105.13544}{{\ttfamily 2105.13544}}].

\bibitem{Wright:2025xka}
A.H.~Wright et~al., \emph{{KiDS-Legacy: Cosmological constraints from cosmic
  shear with the complete Kilo-Degree Survey}},
  \href{https://doi.org/10.1051/0004-6361/202554908}{\emph{Astron. Astrophys.}
  {\bfseries 703} (2025) A158}
  [\href{https://arxiv.org/abs/2503.19441}{{\ttfamily 2503.19441}}].

\bibitem{Knox:2019rjx}
L.~Knox and M.~Millea, \emph{{Hubble constant hunter{\textquoteright}s guide}},
  \href{https://doi.org/10.1103/PhysRevD.101.043533}{\emph{Phys. Rev. D}
  {\bfseries 101} (2020) 043533}
  [\href{https://arxiv.org/abs/1908.03663}{{\ttfamily 1908.03663}}].

\bibitem{Schoneberg:2021qvd}
N.~Sch{\"o}neberg, G.~Franco~Abell{\'a}n, A.~P{\'e}rez~S{\'a}nchez, S.J.~Witte,
  V.~Poulin and J.~Lesgourgues, \emph{{The H0 Olympics: A fair ranking of
  proposed models}},
  \href{https://doi.org/10.1016/j.physrep.2022.07.001}{\emph{Phys. Rept.}
  {\bfseries 984} (2022) 1} [\href{https://arxiv.org/abs/2107.10291}{{\ttfamily
  2107.10291}}].

\bibitem{Guo:2018ans}
R.-Y.~Guo, J.-F.~Zhang and X.~Zhang, \emph{{Can the $H_0$ tension be resolved
  in extensions to $\Lambda$CDM cosmology?}},
  \href{https://doi.org/10.1088/1475-7516/2019/02/054}{\emph{JCAP} {\bfseries
  02} (2019) 054} [\href{https://arxiv.org/abs/1809.02340}{{\ttfamily
  1809.02340}}].

\bibitem{Bernal:2016gxb}
J.L.~Bernal, L.~Verde and A.G.~Riess, \emph{{The trouble with $H_0$}},
  \href{https://doi.org/10.1088/1475-7516/2016/10/019}{\emph{JCAP} {\bfseries
  10} (2016) 019} [\href{https://arxiv.org/abs/1607.05617}{{\ttfamily
  1607.05617}}].

\bibitem{Zhao:2017urm}
M.-M.~Zhao, D.-Z.~He, J.-F.~Zhang and X.~Zhang, \emph{{Search for sterile
  neutrinos in holographic dark energy cosmology: Reconciling Planck
  observation with the local measurement of the Hubble constant}},
  \href{https://doi.org/10.1103/PhysRevD.96.043520}{\emph{Phys. Rev. D}
  {\bfseries 96} (2017) 043520}
  [\href{https://arxiv.org/abs/1703.08456}{{\ttfamily 1703.08456}}].

\bibitem{Feng:2017usu}
L.~Feng, J.-F.~Zhang and X.~Zhang, \emph{{Search for sterile neutrinos in a
  universe of vacuum energy interacting with cold dark matter}},
  \href{https://doi.org/10.1016/j.dark.2018.100261}{\emph{Phys. Dark Univ.}
  {\bfseries 23} (2019) 100261}
  [\href{https://arxiv.org/abs/1712.03148}{{\ttfamily 1712.03148}}].

\bibitem{Blinov:2020hmc}
N.~Blinov and G.~Marques-Tavares, \emph{{Interacting radiation after Planck and
  its implications for the Hubble Tension}},
  \href{https://doi.org/10.1088/1475-7516/2020/09/029}{\emph{JCAP} {\bfseries
  09} (2020) 029} [\href{https://arxiv.org/abs/2003.08387}{{\ttfamily
  2003.08387}}].

\bibitem{Aloni:2021eaq}
D.~Aloni, A.~Berlin, M.~Joseph, M.~Schmaltz and N.~Weiner, \emph{{A Step in
  understanding the Hubble tension}},
  \href{https://doi.org/10.1103/PhysRevD.105.123516}{\emph{Phys. Rev. D}
  {\bfseries 105} (2022) 123516}
  [\href{https://arxiv.org/abs/2111.00014}{{\ttfamily 2111.00014}}].

\bibitem{Poulin:2018cxd}
V.~Poulin, T.L.~Smith, T.~Karwal and M.~Kamionkowski, \emph{{Early Dark Energy
  Can Resolve The Hubble Tension}},
  \href{https://doi.org/10.1103/PhysRevLett.122.221301}{\emph{Phys. Rev. Lett.}
  {\bfseries 122} (2019) 221301}
  [\href{https://arxiv.org/abs/1811.04083}{{\ttfamily 1811.04083}}].

\bibitem{Smith:2019ihp}
T.L.~Smith, V.~Poulin and M.A.~Amin, \emph{{Oscillating scalar fields and the
  Hubble tension: a resolution with novel signatures}},
  \href{https://doi.org/10.1103/PhysRevD.101.063523}{\emph{Phys. Rev. D}
  {\bfseries 101} (2020) 063523}
  [\href{https://arxiv.org/abs/1908.06995}{{\ttfamily 1908.06995}}].

\bibitem{Poulin:2023lkg}
V.~Poulin, T.L.~Smith and T.~Karwal, \emph{{The Ups and Downs of Early Dark
  Energy solutions to the Hubble tension: A review of models, hints and
  constraints circa 2023}},
  \href{https://doi.org/10.1016/j.dark.2023.101348}{\emph{Phys. Dark Univ.}
  {\bfseries 42} (2023) 101348}
  [\href{https://arxiv.org/abs/2302.09032}{{\ttfamily 2302.09032}}].

\bibitem{Hill:2020osr}
J.C.~Hill, E.~McDonough, M.W.~Toomey and S.~Alexander, \emph{{Early dark energy
  does not restore cosmological concordance}},
  \href{https://doi.org/10.1103/PhysRevD.102.043507}{\emph{Phys. Rev. D}
  {\bfseries 102} (2020) 043507}
  [\href{https://arxiv.org/abs/2003.07355}{{\ttfamily 2003.07355}}].

\bibitem{Ivanov:2020ril}
M.M.~Ivanov, E.~McDonough, J.C.~Hill, M.~Simonovi{\'c}, M.W.~Toomey,
  S.~Alexander et~al., \emph{{Constraining Early Dark Energy with Large-Scale
  Structure}}, \href{https://doi.org/10.1103/PhysRevD.102.103502}{\emph{Phys.
  Rev. D} {\bfseries 102} (2020) 103502}
  [\href{https://arxiv.org/abs/2006.11235}{{\ttfamily 2006.11235}}].

\bibitem{Allali:2021azp}
I.J.~Allali, M.P.~Hertzberg and F.~Rompineve, \emph{{Dark sector to restore
  cosmological concordance}},
  \href{https://doi.org/10.1103/PhysRevD.104.L081303}{\emph{Phys. Rev. D}
  {\bfseries 104} (2021) L081303}
  [\href{https://arxiv.org/abs/2104.12798}{{\ttfamily 2104.12798}}].

\bibitem{Chevallier:2000qy}
M.~Chevallier and D.~Polarski, \emph{{Accelerating universes with scaling dark
  matter}}, \href{https://doi.org/10.1142/S0218271801000822}{\emph{Int. J. Mod.
  Phys. D} {\bfseries 10} (2001) 213}
  [\href{https://arxiv.org/abs/gr-qc/0009008}{{\ttfamily gr-qc/0009008}}].

\bibitem{Linder:2002et}
E.V.~Linder, \emph{{Exploring the expansion history of the universe}},
  \href{https://doi.org/10.1103/PhysRevLett.90.091301}{\emph{Phys. Rev. Lett.}
  {\bfseries 90} (2003) 091301}
  [\href{https://arxiv.org/abs/astro-ph/0208512}{{\ttfamily
  astro-ph/0208512}}].

\bibitem{Wang:2016lxa}
B.~Wang, E.~Abdalla, F.~Atrio-Barandela and D.~Pavon, \emph{{Dark Matter and
  Dark Energy Interactions: Theoretical Challenges, Cosmological Implications
  and Observational Signatures}},
  \href{https://doi.org/10.1088/0034-4885/79/9/096901}{\emph{Rept. Prog. Phys.}
  {\bfseries 79} (2016) 096901}
  [\href{https://arxiv.org/abs/1603.08299}{{\ttfamily 1603.08299}}].

\bibitem{DiValentino:2019ffd}
E.~Di~Valentino, A.~Melchiorri, O.~Mena and S.~Vagnozzi, \emph{{Interacting
  dark energy in the early 2020s: A promising solution to the $H_0$ and cosmic
  shear tensions}},
  \href{https://doi.org/10.1016/j.dark.2020.100666}{\emph{Phys. Dark Univ.}
  {\bfseries 30} (2020) 100666}
  [\href{https://arxiv.org/abs/1908.04281}{{\ttfamily 1908.04281}}].

\bibitem{Li:2024qso}
T.-N.~Li, P.-J.~Wu, G.-H.~Du, S.-J.~Jin, H.-L.~Li, J.-F.~Zhang et~al.,
  \emph{{Constraints on Interacting Dark Energy Models from the DESI Baryon
  Acoustic Oscillation and DES Supernovae Data}},
  \href{https://doi.org/10.3847/1538-4357/ad87f0}{\emph{Astrophys. J.}
  {\bfseries 976} (2024) 1} [\href{https://arxiv.org/abs/2407.14934}{{\ttfamily
  2407.14934}}].

\bibitem{Li:2025owk}
T.-N.~Li, G.-H.~Du, Y.-H.~Li, P.-J.~Wu, S.-J.~Jin, J.-F.~Zhang et~al.,
  \emph{{Probing the sign-changeable interaction between dark energy and dark
  matter with DESI baryon acoustic oscillations and DES supernovae data}},
  \href{https://doi.org/10.1007/s11433-025-2771-5}{\emph{Sci. China Phys. Mech.
  Astron.} {\bfseries 69} (2026) 210413}
  [\href{https://arxiv.org/abs/2501.07361}{{\ttfamily 2501.07361}}].

\bibitem{Li:2026xaz}
T.-N.~Li, W.~Giar{\`e}, G.-H.~Du, Y.-H.~Li, E.~Di~Valentino, J.-F.~Zhang
  et~al., \emph{{Robust Preference for Dark Sector Interactions}},
  \href{https://arxiv.org/abs/2601.07361}{{\ttfamily 2601.07361}}.

\bibitem{Li:2025muv}
T.-N.~Li, G.-H.~Du, Y.-H.~Li, Y.~Li, J.-L.~Ling, J.-F.~Zhang et~al.,
  \emph{{Updated constraints on interacting dark energy: A comprehensive
  analysis using multiple CMB probes, DESI DR2, and supernovae observations}},
  \href{https://arxiv.org/abs/2510.11363}{{\ttfamily 2510.11363}}.

\bibitem{Li:2025vuh}
T.-N.~Li, G.-H.~Du, S.-H.~Zhou, Y.-H.~Li, J.-F.~Zhang and X.~Zhang,
  \emph{{Robust evidence for dynamical dark energy in light of DESI DR2 and
  joint ACT, SPT, and Planck data}},
  \href{https://doi.org/10.1016/j.dark.2026.102254}{\emph{Phys. Dark Univ.}
  {\bfseries 52} (2026) 102254}
  [\href{https://arxiv.org/abs/2511.22512}{{\ttfamily 2511.22512}}].

\bibitem{Zhang:2026sci}
X.~Zhang, \emph{{Diagnosing dark energy in the post-DESI era: From effective
  dynamics to multi-probe consistency tests}},
  \href{https://doi.org/10.1007/s11433-026-3116-4}{\emph{Sci. China Phys. Mech.
  Astron.} {\bfseries 69} (2026) 120401}.

\bibitem{Aylor:2018drw}
K.~Aylor, M.~Joy, L.~Knox, M.~Millea, S.~Raghunathan and W.L.K.~Wu,
  \emph{{Sounds Discordant: Classical Distance Ladder {\&} $\Lambda$CDM -based
  Determinations of the Cosmological Sound Horizon}},
  \href{https://doi.org/10.3847/1538-4357/ab0898}{\emph{Astrophys. J.}
  {\bfseries 874} (2019) 4} [\href{https://arxiv.org/abs/1811.00537}{{\ttfamily
  1811.00537}}].

\bibitem{Pourtsidou:2016ico}
A.~Pourtsidou and T.~Tram, \emph{{Reconciling CMB and structure growth
  measurements with dark energy interactions}},
  \href{https://doi.org/10.1103/PhysRevD.94.043518}{\emph{Phys. Rev. D}
  {\bfseries 94} (2016) 043518}
  [\href{https://arxiv.org/abs/1604.04222}{{\ttfamily 1604.04222}}].

\bibitem{Lucca:2021dxo}
M.~Lucca, \emph{{Dark energy{\textendash}dark matter interactions as a solution
  to the S8 tension}},
  \href{https://doi.org/10.1016/j.dark.2021.100899}{\emph{Phys. Dark Univ.}
  {\bfseries 34} (2021) 100899}
  [\href{https://arxiv.org/abs/2105.09249}{{\ttfamily 2105.09249}}].

\bibitem{McDonough:2021pdg}
E.~McDonough, M.-X.~Lin, J.C.~Hill, W.~Hu and S.~Zhou, \emph{{Early dark
  sector, the Hubble tension, and the swampland}},
  \href{https://doi.org/10.1103/PhysRevD.106.043525}{\emph{Phys. Rev. D}
  {\bfseries 106} (2022) 043525}
  [\href{https://arxiv.org/abs/2112.09128}{{\ttfamily 2112.09128}}].

\bibitem{Weinberg:2009wa}
S.~Weinberg, \emph{{Asymptotically Safe Inflation}},
  \href{https://doi.org/10.1103/PhysRevD.81.083535}{\emph{Phys. Rev. D}
  {\bfseries 81} (2010) 083535}
  [\href{https://arxiv.org/abs/0911.3165}{{\ttfamily 0911.3165}}].

\bibitem{Parker:1973qd}
L.~Parker and S.A.~Fulling, \emph{{Quantized matter fields and the avoidance of
  singularities in general relativity}},
  \href{https://doi.org/10.1103/PhysRevD.7.2357}{\emph{Phys. Rev. D} {\bfseries
  7} (1973) 2357}.

\bibitem{Xue:2014kna}
S.-S.~Xue, \emph{{How universe evolves with cosmological and gravitational
  constants}},
  \href{https://doi.org/10.1016/j.nuclphysb.2015.05.022}{\emph{Nucl. Phys. B}
  {\bfseries 897} (2015) 326}
  [\href{https://arxiv.org/abs/1410.6152}{{\ttfamily 1410.6152}}].

\bibitem{Xue:2023qft}
S.-S.~Xue, \emph{{Holographic massive plasma state in Friedman universe:
  cosmological fine-tuning and coincidence problems}},
  \href{https://doi.org/10.1088/1475-7516/2024/05/113}{\emph{JCAP} {\bfseries
  05} (2024) 113} [\href{https://arxiv.org/abs/2309.15488}{{\ttfamily
  2309.15488}}].

\bibitem{Gao:2021xnk}
L.-Y.~Gao, Z.-W.~Zhao, S.-S.~Xue and X.~Zhang, \emph{{Relieving the H 0 tension
  with a new interacting dark energy model}},
  \href{https://doi.org/10.1088/1475-7516/2021/07/005}{\emph{JCAP} {\bfseries
  07} (2021) 005} [\href{https://arxiv.org/abs/2101.10714}{{\ttfamily
  2101.10714}}].

\bibitem{Gao:2022ahg}
L.-Y.~Gao, S.-S.~Xue and X.~Zhang, \emph{{Dark energy and matter interacting
  scenario to relieve H $_{0}$ and S $_{8}$ tensions*}},
  \href{https://doi.org/10.1088/1674-1137/ad2b52}{\emph{Chin. Phys. C}
  {\bfseries 48} (2024) 051001}
  [\href{https://arxiv.org/abs/2212.13146}{{\ttfamily 2212.13146}}].

\bibitem{Zhang:2025dwu}
Y.-M.~Zhang, T.-N.~Li, G.-H.~Du, S.-H.~Zhou, L.-Y.~Gao, J.-F.~Zhang et~al.,
  \emph{{Alleviating the $H_0$ tension through the interacting dark energy
  model from quantum gravitational field theory in light of DESI DR2}},
  \href{https://arxiv.org/abs/2510.12627}{{\ttfamily 2510.12627}}.

\bibitem{Wands:2012vg}
D.~Wands, J.~De-Santiago and Y.~Wang, \emph{{Inhomogeneous vacuum energy}},
  \href{https://doi.org/10.1088/0264-9381/29/14/145017}{\emph{Class. Quant.
  Grav.} {\bfseries 29} (2012) 145017}
  [\href{https://arxiv.org/abs/1203.6776}{{\ttfamily 1203.6776}}].

\bibitem{Wang:2014xca}
Y.~Wang, D.~Wands, G.-B.~Zhao and L.~Xu, \emph{{Post-$Planck$ constraints on
  interacting vacuum energy}},
  \href{https://doi.org/10.1103/PhysRevD.90.023502}{\emph{Phys. Rev. D}
  {\bfseries 90} (2014) 023502}
  [\href{https://arxiv.org/abs/1404.5706}{{\ttfamily 1404.5706}}].

\bibitem{Martinelli:2019dau}
M.~Martinelli, N.B.~Hogg, S.~Peirone, M.~Bruni and D.~Wands, \emph{{Constraints
  on the interacting vacuum{\textendash}geodesic CDM scenario}},
  \href{https://doi.org/10.1093/mnras/stz1915}{\emph{Mon. Not. Roy. Astron.
  Soc.} {\bfseries 488} (2019) 3423}
  [\href{https://arxiv.org/abs/1902.10694}{{\ttfamily 1902.10694}}].

\bibitem{Hogg:2020rdp}
N.B.~Hogg, M.~Bruni, R.~Crittenden, M.~Martinelli and S.~Peirone, \emph{{Latest
  evidence for a late time vacuum{\textendash}geodesic CDM interaction}},
  \href{https://doi.org/10.1016/j.dark.2020.100583}{\emph{Phys. Dark Univ.}
  {\bfseries 29} (2020) 100583}
  [\href{https://arxiv.org/abs/2002.10449}{{\ttfamily 2002.10449}}].

\bibitem{Borges:2017jvi}
H.A.~Borges and D.~Wands, \emph{{Growth of structure in interacting vacuum
  cosmologies}}, \href{https://doi.org/10.1103/PhysRevD.101.103519}{\emph{Phys.
  Rev. D} {\bfseries 101} (2020) 103519}
  [\href{https://arxiv.org/abs/1709.08933}{{\ttfamily 1709.08933}}].

\bibitem{Planck:2019nip}
{\scshape Planck} collaboration, \emph{{Planck 2018 results. V. CMB power
  spectra and likelihoods}},
  \href{https://doi.org/10.1051/0004-6361/201936386}{\emph{Astron. Astrophys.}
  {\bfseries 641} (2020) A5}
  [\href{https://arxiv.org/abs/1907.12875}{{\ttfamily 1907.12875}}].

\bibitem{Planck:2018lbu}
{\scshape Planck} collaboration, \emph{{Planck 2018 results. VIII.
  Gravitational lensing}},
  \href{https://doi.org/10.1051/0004-6361/201833886}{\emph{Astron. Astrophys.}
  {\bfseries 641} (2020) A8}
  [\href{https://arxiv.org/abs/1807.06210}{{\ttfamily 1807.06210}}].

\bibitem{DESI:2025zpo}
{\scshape DESI} collaboration, \emph{{DESI DR2 results. I. Baryon acoustic
  oscillations from the Lyman alpha forest}},
  \href{https://doi.org/10.1103/2wwn-xjm5}{\emph{Phys. Rev. D} {\bfseries 112}
  (2025) 083514} [\href{https://arxiv.org/abs/2503.14739}{{\ttfamily
  2503.14739}}].

\bibitem{Beutler:2011hx}
F.~Beutler, C.~Blake, M.~Colless, D.H.~Jones, L.~Staveley-Smith, L.~Campbell
  et~al., \emph{{The 6dF Galaxy Survey: Baryon Acoustic Oscillations and the
  Local Hubble Constant}},
  \href{https://doi.org/10.1111/j.1365-2966.2011.19250.x}{\emph{Mon. Not. Roy.
  Astron. Soc.} {\bfseries 416} (2011) 3017}
  [\href{https://arxiv.org/abs/1106.3366}{{\ttfamily 1106.3366}}].

\bibitem{Ross:2014qpa}
A.J.~Ross, L.~Samushia, C.~Howlett, W.J.~Percival, A.~Burden and M.~Manera,
  \emph{{The clustering of the SDSS DR7 main Galaxy sample {\textendash} I. A 4
  per cent distance measure at $z = 0.15$}},
  \href{https://doi.org/10.1093/mnras/stv154}{\emph{Mon. Not. Roy. Astron.
  Soc.} {\bfseries 449} (2015) 835}
  [\href{https://arxiv.org/abs/1409.3242}{{\ttfamily 1409.3242}}].

\bibitem{Xue:2021jyj}
S.-S.~Xue, \emph{{Massive particle pair production and oscillation in Friedman
  Universe: its effect on inflation}},
  \href{https://doi.org/10.1140/epjc/s10052-023-11195-6}{\emph{Eur. Phys. J. C}
  {\bfseries 83} (2023) 36} [\href{https://arxiv.org/abs/2112.09661}{{\ttfamily
  2112.09661}}].

\bibitem{Xue:2020tpf}
S.-S.~Xue, \emph{{Massive particle pair production and oscillation in Friedman
  Universe: reheating energy and entropy, and cold dark matter}},
  \href{https://doi.org/10.1140/epjc/s10052-023-11524-9}{\emph{Eur. Phys. J. C}
  {\bfseries 83} (2023) 355}
  [\href{https://arxiv.org/abs/2006.15622}{{\ttfamily 2006.15622}}].

\bibitem{Sanchez:2020vvb}
A.G.~Sanchez, \emph{{Arguments against using $h^{-1}{\rm Mpc}$ units in
  observational cosmology}},
  \href{https://doi.org/10.1103/PhysRevD.102.123511}{\emph{Phys. Rev. D}
  {\bfseries 102} (2020) 123511}
  [\href{https://arxiv.org/abs/2002.07829}{{\ttfamily 2002.07829}}].

\bibitem{Li:2014eha}
Y.-H.~Li, J.-F.~Zhang and X.~Zhang, \emph{{Parametrized Post-Friedmann
  Framework for Interacting Dark Energy}},
  \href{https://doi.org/10.1103/PhysRevD.90.063005}{\emph{Phys. Rev. D}
  {\bfseries 90} (2014) 063005}
  [\href{https://arxiv.org/abs/1404.5220}{{\ttfamily 1404.5220}}].

\bibitem{Ma:1995ey}
C.-P.~Ma and E.~Bertschinger, \emph{{Cosmological perturbation theory in the
  synchronous and conformal Newtonian gauges}},
  \href{https://doi.org/10.1086/176550}{\emph{Astrophys. J.} {\bfseries 455}
  (1995) 7} [\href{https://arxiv.org/abs/astro-ph/9506072}{{\ttfamily
  astro-ph/9506072}}].

\bibitem{Bashinsky:2003tk}
S.~Bashinsky and U.~Seljak, \emph{{Neutrino perturbations in CMB anisotropy and
  matter clustering}},
  \href{https://doi.org/10.1103/PhysRevD.69.083002}{\emph{Phys. Rev. D}
  {\bfseries 69} (2004) 083002}
  [\href{https://arxiv.org/abs/astro-ph/0310198}{{\ttfamily
  astro-ph/0310198}}].

\bibitem{Baumann:2015rya}
D.~Baumann, D.~Green, J.~Meyers and B.~Wallisch, \emph{{Phases of New Physics
  in the CMB}},
  \href{https://doi.org/10.1088/1475-7516/2016/01/007}{\emph{JCAP} {\bfseries
  01} (2016) 007} [\href{https://arxiv.org/abs/1508.06342}{{\ttfamily
  1508.06342}}].

\bibitem{Camarena:2021jlr}
D.~Camarena and V.~Marra, \emph{{On the use of the local prior on the absolute
  magnitude of Type Ia supernovae in cosmological inference}},
  \href{https://doi.org/10.1093/mnras/stab1200}{\emph{Mon. Not. Roy. Astron.
  Soc.} {\bfseries 504} (2021) 5164}
  [\href{https://arxiv.org/abs/2101.08641}{{\ttfamily 2101.08641}}].

\bibitem{Efstathiou:2021ocp}
G.~Efstathiou, \emph{{To H0 or not to H0?}},
  \href{https://doi.org/10.1093/mnras/stab1588}{\emph{Mon. Not. Roy. Astron.
  Soc.} {\bfseries 505} (2021) 3866}
  [\href{https://arxiv.org/abs/2103.08723}{{\ttfamily 2103.08723}}].

\bibitem{Kamionkowski:2014zda}
M.~Kamionkowski, J.~Pradler and D.G.E.~Walker, \emph{{Dark energy from the
  string axiverse}},
  \href{https://doi.org/10.1103/PhysRevLett.113.251302}{\emph{Phys. Rev. Lett.}
  {\bfseries 113} (2014) 251302}
  [\href{https://arxiv.org/abs/1409.0549}{{\ttfamily 1409.0549}}].

\bibitem{Murgia:2020ryi}
R.~Murgia, G.F.~Abell{\'a}n and V.~Poulin, \emph{{Early dark energy resolution
  to the Hubble tension in light of weak lensing surveys and lensing
  anomalies}}, \href{https://doi.org/10.1103/PhysRevD.103.063502}{\emph{Phys.
  Rev. D} {\bfseries 103} (2021) 063502}
  [\href{https://arxiv.org/abs/2009.10733}{{\ttfamily 2009.10733}}].

\bibitem{Smith:2020rxx}
T.L.~Smith, V.~Poulin, J.L.~Bernal, K.K.~Boddy, M.~Kamionkowski and R.~Murgia,
  \emph{{Early dark energy is not excluded by current large-scale structure
  data}}, \href{https://doi.org/10.1103/PhysRevD.103.123542}{\emph{Phys. Rev.
  D} {\bfseries 103} (2021) 123542}
  [\href{https://arxiv.org/abs/2009.10740}{{\ttfamily 2009.10740}}].

\bibitem{Lesgourgues:2011re}
J.~Lesgourgues, \emph{{The Cosmic Linear Anisotropy Solving System (CLASS) I:
  Overview}},  \href{https://arxiv.org/abs/1104.2932}{{\ttfamily 1104.2932}}.

\bibitem{Blas:2011rf}
D.~Blas, J.~Lesgourgues and T.~Tram, \emph{{The Cosmic Linear Anisotropy
  Solving System (CLASS) II: Approximation schemes}},
  \href{https://doi.org/10.1088/1475-7516/2011/07/034}{\emph{JCAP} {\bfseries
  07} (2011) 034} [\href{https://arxiv.org/abs/1104.2933}{{\ttfamily
  1104.2933}}].

\bibitem{Poulin:2018dzj}
V.~Poulin, T.L.~Smith, D.~Grin, T.~Karwal and M.~Kamionkowski,
  \emph{{Cosmological implications of ultralight axionlike fields}},
  \href{https://doi.org/10.1103/PhysRevD.98.083525}{\emph{Phys. Rev. D}
  {\bfseries 98} (2018) 083525}
  [\href{https://arxiv.org/abs/1806.10608}{{\ttfamily 1806.10608}}].

\bibitem{Mead:2020vgs}
A.~Mead, S.~Brieden, T.~Tr{\"o}ster and C.~Heymans, \emph{{hmcode-2020:
  improved modelling of non-linear cosmological power spectra with baryonic
  feedback}}, \href{https://doi.org/10.1093/mnras/stab082}{\emph{Mon. Not. Roy.
  Astron. Soc.} {\bfseries 502} (2021) 1401}
  [\href{https://arxiv.org/abs/2009.01858}{{\ttfamily 2009.01858}}].

\bibitem{Audren:2012wb}
B.~Audren, J.~Lesgourgues, K.~Benabed and S.~Prunet, \emph{{Conservative
  Constraints on Early Cosmology: an illustration of the Monte Python
  cosmological parameter inference code}},
  \href{https://doi.org/10.1088/1475-7516/2013/02/001}{\emph{JCAP} {\bfseries
  02} (2013) 001} [\href{https://arxiv.org/abs/1210.7183}{{\ttfamily
  1210.7183}}].

\bibitem{Brinckmann:2018cvx}
T.~Brinckmann and J.~Lesgourgues, \emph{{MontePython 3: boosted MCMC sampler
  and other features}},
  \href{https://doi.org/10.1016/j.dark.2018.100260}{\emph{Phys. Dark Univ.}
  {\bfseries 24} (2019) 100260}
  [\href{https://arxiv.org/abs/1804.07261}{{\ttfamily 1804.07261}}].

\bibitem{Gelman:1992zz}
A.~Gelman and D.B.~Rubin, \emph{{Inference from Iterative Simulation Using
  Multiple Sequences}},
  \href{https://doi.org/10.1214/ss/1177011136}{\emph{Statist. Sci.} {\bfseries
  7} (1992) 457}.

\bibitem{Akaike:1974vps}
H.~Akaike, \emph{{A new look at the statistical model identification}},
  \href{https://doi.org/10.1109/TAC.1974.1100705}{\emph{IEEE Trans. Automatic
  Control} {\bfseries 19} (1974) 716}.

\bibitem{Raveri:2021wfz}
M.~Raveri and C.~Doux, \emph{{Non-Gaussian estimates of tensions in
  cosmological parameters}},
  \href{https://doi.org/10.1103/PhysRevD.104.043504}{\emph{Phys. Rev. D}
  {\bfseries 104} (2021) 043504}
  [\href{https://arxiv.org/abs/2105.03324}{{\ttfamily 2105.03324}}].

\bibitem{Beutler:2012px}
F.~Beutler, C.~Blake, M.~Colless, D.H.~Jones, L.~Staveley-Smith, G.B.~Poole
  et~al., \emph{{The 6dF Galaxy Survey: $z \approx 0$ measurement of the growth
  rate and $\sigma_8$}},
  \href{https://doi.org/10.1111/j.1365-2966.2012.21136.x}{\emph{Mon. Not. Roy.
  Astron. Soc.} {\bfseries 423} (2012) 3430}
  [\href{https://arxiv.org/abs/1204.4725}{{\ttfamily 1204.4725}}].

\bibitem{Howlett:2014opa}
C.~Howlett, A.~Ross, L.~Samushia, W.~Percival and M.~Manera, \emph{{The
  clustering of the SDSS main galaxy sample {\textendash} II. Mock galaxy
  catalogues and a measurement of the growth of structure from redshift space
  distortions at $z = 0.15$}},
  \href{https://doi.org/10.1093/mnras/stu2693}{\emph{Mon. Not. Roy. Astron.
  Soc.} {\bfseries 449} (2015) 848}
  [\href{https://arxiv.org/abs/1409.3238}{{\ttfamily 1409.3238}}].

\bibitem{eBOSS:2020yzd}
{\scshape eBOSS} collaboration, \emph{{Completed SDSS-IV extended Baryon
  Oscillation Spectroscopic Survey: Cosmological implications from two decades
  of spectroscopic surveys at the Apache Point Observatory}},
  \href{https://doi.org/10.1103/PhysRevD.103.083533}{\emph{Phys. Rev. D}
  {\bfseries 103} (2021) 083533}
  [\href{https://arxiv.org/abs/2007.08991}{{\ttfamily 2007.08991}}].

\bibitem{Reischke:2024fvk}
R.~Reischke et~al., \emph{{KiDS-Legacy: Covariance validation and the unified
  OneCovariance framework for projected large-scale structure observables}},
  \href{https://doi.org/10.1051/0004-6361/202452592}{\emph{Astron. Astrophys.}
  {\bfseries 699} (2025) A124}
  [\href{https://arxiv.org/abs/2410.06962}{{\ttfamily 2410.06962}}].

\bibitem{LoVerde:2008re}
M.~LoVerde and N.~Afshordi, \emph{{Extended Limber Approximation}},
  \href{https://doi.org/10.1103/PhysRevD.78.123506}{\emph{Phys. Rev. D}
  {\bfseries 78} (2008) 123506}
  [\href{https://arxiv.org/abs/0809.5112}{{\ttfamily 0809.5112}}].

\bibitem{Hirata:2004gc}
C.M.~Hirata and U.~Seljak, \emph{{Intrinsic alignment-lensing interference as a
  contaminant of cosmic shear}},
  \href{https://doi.org/10.1103/PhysRevD.82.049901}{\emph{Phys. Rev. D}
  {\bfseries 70} (2004) 063526}
  [\href{https://arxiv.org/abs/astro-ph/0406275}{{\ttfamily
  astro-ph/0406275}}].

\bibitem{Bridle:2007ft}
S.~Bridle and L.~King, \emph{{Dark energy constraints from cosmic shear power
  spectra: impact of intrinsic alignments on photometric redshift
  requirements}}, \href{https://doi.org/10.1088/1367-2630/9/12/444}{\emph{New
  J. Phys.} {\bfseries 9} (2007) 444}
  [\href{https://arxiv.org/abs/0705.0166}{{\ttfamily 0705.0166}}].

\bibitem{Wu:2025vrl}
P.-J.~Wu, M.~Zhang and S.-J.~Jin, \emph{{How dark sector equations of state
  govern interaction signatures}},
  \href{https://doi.org/10.1103/zg3k-k6g1}{\emph{Phys. Rev. D} {\bfseries 113}
  (2026) L081302} [\href{https://arxiv.org/abs/2512.05548}{{\ttfamily
  2512.05548}}].

\bibitem{AtacamaCosmologyTelescope:2025nti}
{\scshape Atacama Cosmology Telescope} collaboration, \emph{{The Atacama
  Cosmology Telescope: DR6 constraints on extended cosmological models}},
  \href{https://doi.org/10.1088/1475-7516/2025/11/063}{\emph{JCAP} {\bfseries
  11} (2025) 063} [\href{https://arxiv.org/abs/2503.14454}{{\ttfamily
  2503.14454}}].

\bibitem{Euclid:2024yrr}
{\scshape Euclid} collaboration, \emph{{Euclid - I. Overview of the Euclid
  mission}}, \href{https://doi.org/10.1051/0004-6361/202450810}{\emph{Astron.
  Astrophys.} {\bfseries 697} (2025) A1}
  [\href{https://arxiv.org/abs/2405.13491}{{\ttfamily 2405.13491}}].

\bibitem{LSST:2008ijt}
{\scshape LSST} collaboration, \emph{{LSST: from Science Drivers to Reference
  Design and Anticipated Data Products}},
  \href{https://doi.org/10.3847/1538-4357/ab042c}{\emph{Astrophys. J.}
  {\bfseries 873} (2019) 111}
  [\href{https://arxiv.org/abs/0805.2366}{{\ttfamily 0805.2366}}].

\bibitem{SKA:2018ckk}
{\scshape SKA} collaboration, \emph{{Cosmology with Phase 1 of the Square
  Kilometre Array: Red Book 2018: Technical specifications and performance
  forecasts}}, \href{https://doi.org/10.1017/pasa.2019.51}{\emph{Publ. Astron.
  Soc. Austral.} {\bfseries 37} (2020) e007}
  [\href{https://arxiv.org/abs/1811.02743}{{\ttfamily 1811.02743}}].

\bibitem{Bull:2014rha}
P.~Bull, P.G.~Ferreira, P.~Patel and M.G.~Santos, \emph{{Late-time cosmology
  with 21cm intensity mapping experiments}},
  \href{https://doi.org/10.1088/0004-637X/803/1/21}{\emph{Astrophys. J.}
  {\bfseries 803} (2015) 21} [\href{https://arxiv.org/abs/1405.1452}{{\ttfamily
  1405.1452}}].

\bibitem{Wu:2021vfz}
P.-J.~Wu and X.~Zhang, \emph{{Prospects for measuring dark energy with 21 cm
  intensity mapping experiments}},
  \href{https://doi.org/10.1088/1475-7516/2022/01/060}{\emph{JCAP} {\bfseries
  01} (2022) 060} [\href{https://arxiv.org/abs/2108.03552}{{\ttfamily
  2108.03552}}].

\bibitem{Zhang:2021yof}
M.~Zhang, B.~Wang, P.-J.~Wu, J.-Z.~Qi, Y.~Xu, J.-F.~Zhang et~al.,
  \emph{{Prospects for Constraining Interacting Dark Energy Models with 21 cm
  Intensity Mapping Experiments}},
  \href{https://doi.org/10.3847/1538-4357/ac0ef5}{\emph{Astrophys. J.}
  {\bfseries 918} (2021) 56}
  [\href{https://arxiv.org/abs/2102.03979}{{\ttfamily 2102.03979}}].

\end{thebibliography}\endgroup

\end{document}